\documentclass[amsfonts,amsmath,floatfix,twocolumn]{revtex4}

\usepackage{graphicx}
\usepackage{epsfig}
\usepackage{epstopdf}
\usepackage[usenames,dvipsnames]{xcolor}
\usepackage{braket}
\usepackage{hyperref}
\usepackage{amsthm}
\usepackage{amssymb}
\usepackage{enumitem}
\usepackage{dsfont}
\usepackage{bbold}
\usepackage{eurosym}
\usepackage[normalem]{ulem}
\usepackage{soul}

\newcommand{\be}{\begin{equation}}
\newcommand{\ee}{\end{equation}}
\newcommand{\ba}{\begin{aligned}}
\newcommand{\ea}{\end{aligned}}

\newcommand{\bc}{\begin{center}}
\newcommand{\ec}{\end{center}}
\newcommand{\beq}{\begin{equation}}
\newcommand{\eeq}{\end{equation}}
\newcommand{\beqq}{\begin{equation*}}
\newcommand{\eeqq}{\end{equation*}}
\newcommand{\beqa}{\begin{align}}
\newcommand{\eeqa}{\end{align}}
\newcommand{\barr}{\begin{array}}
\newcommand{\earr}{\end{array}}
\newcommand{\bi}{\begin{itemize}}
\newcommand{\ei}{\end{itemize}}

\newcommand{\Hi}{\mathcal{H}}

\theoremstyle{remark}
\newtheorem{lem}{Lemma}
\newtheorem{theo}{Theorem}

\newtheorem{defi}{Definition}

\begin{document}

\title{Stellar representation of non-Gaussian quantum states}

\author{Ulysse Chabaud$^1$}
\email{ulysse.chabaud@gmail.com}
\author{Damian Markham$^1$}
\author{Fr\'ed\'eric Grosshans$^{2,1}$}
\address{$^1$Laboratoire d'Informatique de Paris 6, CNRS, Sorbonne Universit\'e, 4 place Jussieu, 75005 Paris, France}
\address{$^2$Laboratoire Aim\'e Cotton, CNRS, Univ.\@ Paris-Sud, ENS Cachan, Univ.\@ Paris-Saclay, 91405 Orsay Cedex}

\date{\today}


\begin{abstract}

The so-called stellar formalism allows to represent the non-Gaussian properties of single-mode quantum states by the distribution of the zeros of their Husimi $Q$-function in phase-space. We use this representation in order to derive an infinite hierarchy of single-mode states based on the number of zeros of the Husimi $Q$-function, the \textit{stellar hierarchy}. We give an operational characterisation of the states in this hierarchy with the minimal number of single-photon additions needed to engineer them, and derive equivalence classes under Gaussian unitary operations. We study in detail the topological properties of this hierarchy with respect to the trace norm, and discuss implications for non-Gaussian state engineering, and continuous variable quantum computing.

\end{abstract}


\maketitle


\noindent Quantum information processing takes advantage of non-classical phenomena, such as superposition and entanglement, to provide applications beyond what classical information processing may offer~\cite{bennett1993teleporting,shor1994algorithms}. Quantum information may be encoded in physical systems using either discrete variables, e.g.\@ the polarisation of a photon, or continuous variables, e.g.\@ quadratures of the electromagnetic field. Continuous variable quantum information processing~\cite{Braunstein2005} represents a powerful alternative to its discrete variable counterpart, as deterministic generation of highly entangled states~\cite{yokoyama2013ultra,yoshikawa2016invited} and high efficiency measurement are readily available with current technologies.

In continuous variable quantum information, quantum states are described mathematically by vectors in a separable Hilbert space of infinite dimension. Alternatively, phase-space formalism allows to describe quantum states conveniently using generalised quasi-probability distributions~\cite{cahill1969density}, among which are the Husimi $Q$-function, the Wigner $W$-function, and the Glauber-Sudarshan $P$-function, which is always singular. The states that have a Gaussian Wigner or Husimi function are called Gaussian states, while all the other states are called non-Gaussian. By extension, the operations mapping Gaussian states to Gaussian states are called Gaussian operations, and measurements projecting onto Gaussian states are called Gaussian measurements.
Gaussian states and processes feature an elegant mathematical description with the symplectic formalism, and are useful for a wide variety quantum information protocols~\cite{marian1993squeezed,ferraro2005gaussian,Lloyd2012,adesso2014continuous}. However, Gaussian computations, composed of input Gaussian states, Gaussian operations, and Gaussian measurements, are easy to simulate classically~\cite{mari2012positive}. On the other hand, non-Gaussian states are needed, and are actually useful for achieving universal qubit quantum computing with continuous variables~\cite{Gottesman2001,baragiola2019all}, and they are crucial for many other quantum information tasks~\cite{ghose2007non,eisert2002distilling,fiuravsek2002gaussian,giedke2002characterization,wenger2003maximal,garcia2004proposal,niset2009no,adesso2009optimal,barbosa2019continuous}. Characterizing and understanding the properties of these states is thus of major importance~\cite{takagi2018convex,zhuang2018resource,albarelli2018resource,lami2018gaussian}.

Hudson~\cite{hudson1974wigner} has notably shown that a single-mode pure quantum state is non-Gaussian if and only if its Wigner function has negative values, and this result has been generalised to multimode states by Soto and Claverie~\cite{soto1983wigner}. 
This characterization is an interesting starting point for studying non-Gaussian states. From this result, one can introduce measures of a state being non-Gaussian using Wigner negativity, e.g.\@ the negative volume~\cite{kenfack2004negativity}, that are invariant under Gaussian operations.
However, computing these quantities from experimental data is complicated in practice. 
Other measures and witnesses for non-Gaussian states have been derived~\cite{genoni2007measure,filip2011detecting,genoni2013detecting,hughes2014quantum}, which allow to discriminate non-Gaussian states from mixtures of Gaussian states from experimental data.

The Husimi function, which is a smoothed version of the Wigner function, also allows for characterizing non-Gaussian states: for pure states, the Husimi function having zeros is actually equivalent to the Wigner function having negative values, as shown by L\"utkenhaus and Barnett~\cite{lutkenhaus1995nonclassical}. Informally,

\begin{theo}---\label{Husimi}
A pure quantum state is non-Gaussian if and only if its Husimi $Q$-function has zeros.
\end{theo}

An interesting point is that for single-mode states, the zeros of the Husimi $Q$-function form a discrete set, as we will show later on. The non-Gaussian properties of single-mode states may thus be described by the distribution of these zeros in phase-space.

Based on this result, we introduce in this Letter an infinite hierarchy of states, which we call \textit{stellar hierarchy}, which allows to characterize single-mode continuous variable pure quantum states with respect to their non-Gaussian properties. We make use of the so-called stellar representation, or Segal-Bargmann formalism~\cite{bargmann1961hilbert,segal1963mathematical}, in order to derive this hierarchy. We give a brief introduction to this formalism in what follows, and we review and prove additional relevant properties. We then define the \textit{stellar rank} of a state, which induces the stellar hierarchy, and we characterize the set of states of each rank. In particular, we show that each rank is left invariant under Gaussian operations. At rank zero lie Gaussian states, while non-Gaussian states populate all higher ranks. We show that the stellar rank of a state is equivalent to the minimal number of photon additions necessary to engineer the state. We then use this hierarchy to study analytically Gaussian convertibility of states, and we derive equivalence classes under this relation. We then study the topology of the stellar hierarchy, with respect to the trace norm, and show that it is robust. We show that the stellar hierarchy matches the hierarchy of genuine $n$-photon quantum non-Gaussian light introduced in~\cite{lachman2018faithful}, and we discuss implications of our results for non-Gaussian quantum state engineering, and continuous variable quantum computing.


\textit{The stellar function}.\@ ---\@ The so-called stellar representation, or Segal-Bargmann representation~\cite{bargmann1961hilbert,segal1963mathematical}, has been used to study quantum chaos~\cite{leboeuf1990chaos,arranz1996distribution,korsch1997zeros,biswas1999distribution}, and the completeness of sequences of coherent states~\cite{perelomov1971completeness,bacry1975proof,boon1978discrete}. 
We give hereafter an introduction to this formalism. Further details may be found e.g.\@ in~\cite{vourdas2006analytic}. 

Let $\mathcal H_\infty$ be the infinite-dimensional Hilbert space of single-mode pure quantum states. In the following, we consider normalised states, and we denote by $\{\ket n\}_{n\in\mathbb N}$ the Fock basis of $\mathcal H_\infty$. We introduce below the stellar function. This function has been recently studied, in the context of non-Gaussian quantum state engineering~\cite{PhysRevA.99.053816}, in order to simplify calculations related to photon-subtracted Gaussian states.

\begin{defi}---
Let $\ket\psi=\sum_{n\ge0}{\psi_n\ket n}\in\mathcal H_\infty$ be a normalised state. The \textit{stellar function} of the state $\ket\psi$ is defined as
\be
F_\psi^\star(\alpha)=e^{\frac12|\alpha|^2}\braket{\alpha^*|\psi}=\sum_{n\ge0}{\psi_n\frac{\alpha^n}{\sqrt{n!}}},
\ee
for all $\alpha\in\mathbb C$, where $\ket\alpha=e^{-\frac12|\alpha|^2}\sum_{n\ge0}{\frac{\alpha^n}{\sqrt{n!}}\ket n}\in\mathcal H_\infty$ is the coherent state of amplitude $\alpha$.
\end{defi}

The stellar function is a holomorphic function over the complex plane, which provides an analytic representation of a quantum state.
For any state $\ket\psi\in\Hi_\infty$, we may write
\be
\ket\psi=\sum_{n\ge0}{\psi_n\ket n}=F_\psi^\star(\hat a^\dag)\ket0,
\label{Fadag}
\ee
using the definition of the stellar function.
An important result is that the stellar representation is unique, up to a global phase:

\begin{lem}---\label{unique}
Let $\ket\phi$ and $\ket\psi$ be pure normalised single-mode states such that $F_\phi^\star=F_\psi^\star$, up to a phase. Then $\ket\phi=\ket\psi$.
Moreover, let $\ket\chi=f(\hat a^\dag)\ket0$ be a single-mode normalised pure state, where $f$ is analytical. Then $f=F_\chi^\star$ up to a phase.
\end{lem}

These results follow directly from Eq.~(\ref{Fadag}), as detailed in the Supplemental Material.

The stellar function of a state $\ket\psi\in\Hi_\infty$ is related to its Husimi $Q$-function, a smoothed version of the Wigner function~\cite{cahill1969density}, given by
\be
Q_\psi(\alpha)=\frac1\pi|\braket{\alpha|\psi}|^2=\frac{e^{-|\alpha|^2}}\pi|F_\psi^\star(\alpha^*)|^2,
\ee
for all $\alpha\in\mathbb C$. The zeros of the Husimi $Q$-function are the complex conjugates of the zeros of $F_\psi^\star$. 
Hence, by Theorem~\ref{Husimi}, a single-mode pure quantum state is non-Gaussian if and only if its stellar function has zeros. These zeros form a discrete set, as the stellar function is a non-zero analytical function. 
The non-Gaussian properties of a single-mode pure state are then described by the distribution of the zeros over the complex plane. Using anti-stereographic projection~\cite{sidoli2007arabic}, this amounts to describing the non-Gaussian properties of a pure state with a set of points on the sphere, hence the name stellar representation, where the points on the sphere looked at from the center of the sphere are seen as stars on the celestial vault~\cite{tualle1995normal,korsch1997zeros}.


\textit{The stellar rank}.\@ ---\@ The Hilbert space $\Hi_\infty$ is naturally partitioned into classes of sets having the same number of zeros. We introduce the following related definition:

\begin{defi}---
The \textit{stellar rank} $r^\star(\psi)$ of a pure single-mode normalised quantum state $\ket\psi\in\Hi_\infty$ is defined as the number of zeros of its stellar function $F_\psi^\star$, counted with multiplicity.
\end{defi}

We introduce hereafter the notation $\overline{\mathbb N}=\mathbb N\cup\{+\infty\}$, so that $r^\star(\psi)\in\overline{\mathbb N}$. For $N\in\overline{\mathbb N}$, we define
\be
R_N=\{\ket\psi\in\Hi_\infty,\text{ }r^\star(\psi)=N\}
\label{finiterng}
\ee
the set of states with stellar rank equal to $N$. The \textit{stellar hierarchy} is the hierarchy of states induced by the stellar rank. By Lemma~\ref{unique}, if $M\neq N$ then $R_M\cap R_N=\varnothing$, for all $M,N\in\overline{\mathbb N}$, so all the ranks in the stellar hierarchy are disjoint. We have $\Hi_\infty=\bigcup_{N\in\overline{\mathbb N}}{R_N}$, i.e.\@ the stellar hierarchy covers the whole space of normalised states, and the set of states of finite stellar rank is given by $\bigcup_{N\in\mathbb N}{R_N}$. 
By Theorem~\ref{Husimi}, the rank zero of the stellar hierarchy $R_0$ is the set of single-mode normalised pure Gaussian states.
For all $N\in\mathbb N$ the photon number state $\ket N$ is of stellar rank $N$, since $F^\star_{\ket N}(\alpha)=\frac{\alpha^N}{\sqrt{N!}}$, while the cat state $\ket{\mathrm{cat}}\propto(\ket{ix}-\ket{-ix})$ is of infinite stellar rank, since $F^\star_{\ket{\mathrm{cat}}}(\alpha)\propto\sin(\alpha x)$, so all ranks are non empty.

By analogy with the Schmidt rank in entanglement theory~\cite{terhal2000schmidt}, we define the stellar rank of a mixed state $\rho$ as $r^\star(\rho)=\inf_{p_i,\psi_i}\sup r^\star(\psi_i)$, where the infimum is over the statistical ensembles such that $\rho=\sum_i{p_i\ket{\psi_i}\bra{\psi_i}}$. 

In the following, we investigate further the properties of the stellar hierarchy. We prove a first general decomposition result for pure states of finite stellar rank:

\begin{theo}---\label{finitez}
Let $\ket\psi\in\bigcup_{N\in\mathbb N}{R_N}$ be a pure state of finite stellar rank. Let $\{\beta_1,\dots,\beta_{r^\star(\psi)}\}$ be the roots of the Husimi $Q$-function of $\ket\psi$, counted with multiplicity. Then,
\be
\ket\psi=\frac1{\mathcal N}\left[\prod_{n=1}^{r^\star(\psi)}{\hat D(\beta_n)\hat a^\dag\hat D^\dag(\beta_n)}\right]\ket{G_\psi},
\ee
where $\hat D(\beta)$ is a displacement operator, $\ket{G_\psi}$ is a Gaussian state, and $\mathcal N$ is a normalisation constant. Moreover, this decomposition is unique up to reordering of the roots.
\end{theo}

The proof of this statement, which combines Eq.~(\ref{Fadag}) with the Hadamard--Weierstrass factorization theorem~\cite{saks1952analytic}, is detailed in the Supplemental Material.

This decomposition implies that any state of finite stellar rank may be obtained from Gaussian states by successive applications of the creation operator at different locations in phase-space, given by the zeros of the Husimi $Q$-function. Experimentally, this corresponds to the probabilistic non-Gaussian operation of single-photon addition~\cite{zavatta2004quantum,marco2010manipulating,walschaers2018tailoring}. 
Using this decomposition, we obtain the following property:

\begin{theo}---\label{Ginv}
A unitary operation is Gaussian if and only if it leaves the stellar rank invariant.
\end{theo}

This result follows directly from Theorem~\ref{finitez}, and we refer to the Supplemental Material for a formal proof.

An interesting consequence is that the number of single-photon additions in the decomposition of Theorem~\ref{finitez} is minimal. Indeed, if a quantum state is obtained from the vacuum by successive applications of Gaussian operations and single-photon additions, then its stellar rank is exactly the number of photon additions, because each single-photon addition increases by one its stellar rank (it adds a zero to the stellar function at zero), while each Gaussian operation leaves the stellar rank invariant by Theorem~\ref{Ginv}. Hence, the stellar rank is a measure of the non-Gaussian properties of a quantum state which may be interpreted as a minimal non-Gaussian operational cost, in terms of single-photon additions, for engineering the state from the vacuum.

\textit{Gaussian convertibility}.\@ ---\@ Now that the first properties of the stellar hierarchy are laid out, we consider as an application the convertibility of quantum states using Gaussian unitary operations:

\begin{defi}---\label{Gconvert}
Two states $\ket\phi$ and $\ket\psi$ are \textit{Gaussian-convertible} if there exists a Gaussian unitary operation $\hat G$ such that $\ket\psi=\hat G\ket\phi$.
\end{defi}

Note that this notion is different from the more restrictive notion of Gaussian conversion introduced in~\cite{yadin2018operational}, which denotes the conversion of Gaussian states with passive linear optics, and a subclass of Gaussian measurements and feed-forward.

Gaussian convertibility defines an equivalence relation in $\mathcal H_\infty$. By Theorem~\ref{Ginv}, having the same stellar rank is a necessary condition for Gaussian convertibility. However, this condition is not sufficient. In order to derive the equivalence classes for Gaussian convertibility, we introduce the following definition:

\begin{defi}---
\textit{Core states} are defined as the single-mode normalised pure quantum states which have a polynomial stellar function.
\end{defi}

By Eq.~(\ref{Fadag}) and Lemma~\ref{unique}, core states are the states with a bounded support over the Fock basis, i.e.\@ finite superpositions of Fock states. These correspond to the \textit{minimal non-Gaussian core states} introduced in~\cite{menzies2009gaussian}, in the context of non-Gaussian state engineering. With this definition, we can state our result on Gaussian convertibility of states of finite stellar rank:

\begin{theo}---\label{Gconv}
Let $\ket\psi\in\bigcup_{N\in\mathbb N}{R_N}$ be a state of finite stellar rank. Then, there exists a unique core state $\ket{C_\psi}$ such that $\ket\psi$ and $\ket{C_\psi}$ are Gaussian-convertible.

By Theorem~\ref{finitez}, $\ket\psi=P_\psi(\hat a^\dag)\ket{G_\psi}$, where $P_\psi$ is a polynomial of degree $r^\star(\psi)$ and $\ket{G_\psi}=\hat S(\xi)\hat D(\beta)\ket0$ is a Gaussian state, where $\hat D(\beta)=e^{\beta\hat a^\dag-\beta^*\hat a}$ is a displacement operator, and $\hat S(\xi)=e^{\frac12(\xi\hat a^2-\xi^*\hat a^{\dag2})}$ is a squeezing operator, with $\xi=re^{i\theta}$. Then,
\be
\ket\psi=\hat S(\xi)\hat D(\beta)\ket{C_\psi}=\hat S(\xi)\hat D(\beta)F_{C_\psi}^\star(\hat a^\dag)\ket0,
\label{decomp2}
\ee
where the (polynomial) stellar function of $\ket{C_\psi}$ is given by
\be
F_{C_\psi}^\star(\alpha)=P_\psi\left(c_r\alpha-s_re^{i\theta}\partial_\alpha+c_r\beta^*-s_re^{i\theta}\beta\right)\cdot1,
\label{polycore}
\ee
for all $\alpha\in\mathbb C$.
\end{theo}

The proof of this result follows from combining Theorem~\ref{finitez} together with Lemma~\ref{unique}, and is detailed in the Supplemental Material.

This result has several important consequences. 
\textit{Firstly}, it implies a second general decomposition result, in addition to Theorem~\ref{finitez}: by Eq.~(\ref{decomp2}), any state of finite stellar rank can be uniquely decomposed as a finite superposition of equally displaced and equally squeezed number states. This shows that the stellar hierarchy matches the genuine $n$-photon hierarchy introduced in~\cite{lachman2018faithful}: a pure state exhibits \textit{genuine} $n$\textit{-photon quantum non-Gaussianity} if and only if it has a stellar rank greater or equal \mbox{to $n$}.
Formally, for all $N\in\mathbb N$, the set $R_N$ of states of stellar rank equal to $N$ is obtained by the free action of the group of single-mode Gaussian unitary operations $\mathcal G$ on the set of core states of stellar rank $N$, which is isomorphic to the set of normalised complex polynomials of degree $N$.\\
\textit{Secondly}, it also gives an analytical way to check if two states of finite stellar rank are Gaussian-convertible, given their stellar functions, by checking with Eq.~(\ref{polycore}) if they share the same core state. A simple example is given in the Supplemental Material, where it is shown using this criterion that single photon states and single photon-subtracted squeezed vacuum states are Gaussian-convertible.\\
\textit{Thirdly}, it shows that two different core states are never Gaussian-convertible, while any state of finite stellar rank is always Gaussian-convertible to a unique core state. This implies that equivalence classes for Gaussian convertibility for states of finite stellar rank correspond to the orbits of core states under Gaussian operations.


\textit{Stellar robustness}.\@ ---\@ Having characterized the states of finite stellar rank, we study in the following the topology of the stellar hierarchy, with respect to the trace norm. In order to discuss the robustness of this hierarchy up to small deviation in trace distance, we introduce the following definition:

\begin{defi}---
Let $\ket\psi\in\Hi_\infty$. The \textit{stellar robustness} of the state $\ket\psi$ is defined as
\be
R^\star(\psi)=\inf_{r^\star(\phi)<r^\star(\psi)}{D_1(\phi,\psi)},
\ee
where $D_1$ denotes the trace distance, and where the infimum is over all states $\ket\phi\in\Hi_\infty$ such that $r^\star(\phi)<r^\star(\psi)$ (with the convention $N<+\infty\Leftrightarrow N\in\mathbb N$).
\end{defi}

The stellar robustness quantifies how much one has to deviate from a quantum state in trace distance to find another quantum state of lower stellar rank. A similar notion is the quantum non-Gaussian depth~\cite{straka2018quantum}, which quantifies the maximum attenuation applicable on a quantum state, after which quantum non-Gaussianity can still be witnessed.
The stellar robustness inherits the property of invariance under Gaussian operations of the stellar rank, because the trace distance is invariant under unitary operations. It is related to the fidelity by the following result:

\begin{lem}---
Let $\ket\psi\in\mathcal H_\infty$, then
\begin{equation}
\sup_{r^\star(\rho)<r^\star(\psi)}F(\rho,\psi)=1-[R^\star(\psi)]^2,
\end{equation}
where $F$ is the fidelity.
\end{lem}
We give a proof in the Supplemental Material. Certifying that a (mixed) state $\rho$ has a fidelity greater than $1-[R^\star(\psi)]^2$ with a given target pure state $\ket\psi$ thus ensures that the state $\rho$ has stellar rank equal or greater that $r^\star(\psi)$.

We characterize hereafter the topology of the stellar hierarchy, with respect to the trace norm. Formally, this topology is summarised by the following result for states of finite stellar rank:

\begin{theo}---
For all $N\in\mathbb N$,
\be
\overline{R_N}=\underset{0\le K\le N}{\bigcup}{R_K},
\ee
where $\overline X$ denotes the closure of $X$ for the trace norm in the set of normalised states $\mathcal H_\infty$.
\label{topology}
\end{theo}

The proof of this result, given in the Supplemental Material, is quite technical, and obtained by showing double inclusion, by considering converging sequences of states and studying their limit. 

This result implies that the set on the right-hand-side, containing the states of stellar rank smaller than $N$, is a closed set in $\Hi_\infty$ for the trace norm. In particular, since all ranks of the stellar hierarchy are disjoint, for any state of finite rank $N$, there is no sequence of states of strictly lower rank converging to it, and this holds for all $N$. Each state of a given finite stellar rank is thus isolated from the lower stellar ranks, i.e.\@ there is a ball around it in trace norm which only contains states of equal or higher stellar rank.
On the other hand, with the other inclusion, no state of a given finite stellar rank is isolated from any higher stellar rank, i.e.\@ one can always find a sequence of states of any higher rank converging to this state in trace norm. 
Hence, Theorem~\ref{topology} implies that for all states $\ket\psi\in\bigcup_{N\in\mathbb N}{R_N}$, we have $R^\star(\psi)>0$, i.e.\@ states of finite stellar rank are robust.
We show in the Supplemental Material that the robustness of a single photon-added squeezed state is given by \mbox{$(1-\frac{3\sqrt3}{4e})^{1/2}\approx0.72$} as an example, and we reduce computing the robustness of any finite stellar rank state to a generic optimization problem.

For states of infinite stellar rank, we have the following result: 

\begin{lem}---\label{dense}
The set of states of finite stellar rank is dense for the trace norm in the set of normalised pure single-mode states:
\be
\overline{\underset{N\in\mathbb N}{\bigcup}{R_N}}=\mathcal H_\infty,
\ee
where $\overline X$ denotes the closure of $X$ for the trace norm in the set of normalised states $\mathcal H_\infty$.
\end{lem}

This result is easily proven by considering the sequence of normalised truncated states for any given state in $\Hi_\infty$. We refer to the Supplemental Material for details.

In particular, this means that states of infinite stellar rank are not isolated from lower stellar ranks, unlike states of finite stellar rank.
Lemma~\ref{dense} thus implies that for all states $\ket\psi\in R_\infty$ of infinite stellar rank, $R^\star(\psi)=0$, i.e.\@ states of infinite stellar rank are not robust.

Preparing a state $\ket\psi$ with precision better than $R^\star(\psi)$ ensures that the obtained state has rank equal or greater than $r^\star(\psi)$. 
For example, engineering a state that has a trace distance less than $(1-\frac{3\sqrt3}{4e})^{1/2}\approx0.72$ with any single photon-added squeezed state implies that this state has a stellar greater or equal to $1$.
When considering imperfect single-mode non-Gaussian state engineering, one may thus restrict to states of finite stellar rank, which are obtained uniquely by a finite number of single-photon additions to a Gaussian state, by Theorem~\ref{finitez}. In particular, cat states, being states of infinite stellar rank, can be approximated to arbitrary precision by finite rank states~\cite{wang2013conditional}. Alternatively, one may also describe such states using Theorem~\ref{Gconv} as finite superposition of displaced squeezed number states. Engineering of such states has recently been considered in~\cite{su2019conversion}, by photon detection of Gaussian states.


\textit{Smoothed non-Gaussianity of formation}.\@ ---\@ The topology of the stellar hierarchy obtained previously motivates the following definition:

\begin{defi}---
Let $\rho$ be a single-mode normalised state, and let $\epsilon>0$. The $\epsilon$-smoothed non-Gaussianity of formation $\mathcal N\mathcal G\mathcal F_\epsilon(\rho)$ is defined as the minimal stellar rank of the states $\sigma$ that are $\epsilon$-close to $\rho$ in trace distance. Formally,
\be
\mathcal N\mathcal G\mathcal F_\epsilon(\rho)=\inf_\sigma{\left\{r^\star(\sigma),\text{ s.t. }D_1(\rho,\sigma)\le\epsilon\right\}},
\ee
where $D_1$ denotes the trace distance.
\label{NGformation}
\end{defi}

The infimum is also a minimum, since the set considered only contains integer values and is lower bounded by zero.
That minimum is not necessarily attained for the energy cut-off state (consider e.g.\@ a Gaussian state).
The smoothed non-Gaussianity of formation is a smoothed version of the stellar rank.
By Theorem~\ref{finitez}, it quantifies the minimal number of single-photon additions that need to be applied to a Gaussian state in order to obtain a state $\epsilon$-close to a target state.
As mentioned in the introduction, universal qubit quantum computing with continuous variables may be achieved using specific non-Gaussian resource states together with Gaussian operations and measurements~\cite{baragiola2019all}. On the other hand, the trace distance between two states provides a meaningful measure in the context of quantum computing, because a small trace distance ensures that any computation done with the states will yield similar results, with high probability~\cite{NielsenChuang}. In that context, the smoothed non-Gaussianity of formation provides an operational cost measure for non-Gaussian resource states, which is invariant under Gaussian operations.


\textit{Summary and discussion}.\@ ---\@ Based on the stellar representation of single-mode continuous variable quantum states, we have defined the stellar rank as the number of zeros of the stellar function, or equivalently of the Husimi $Q$-function. Using the analytical properties of the stellar function, we have shown that this rank is invariant under Gaussian operations, and induces a hierarchy over the space of single-mode normalised states. We have characterized the states of finite stellar rank as the states obtained by successive single-photon additions to a Gaussian state, or equivalently as finite superpositions of (equally) displaced and squeezed number states. Additionally, we have given the stellar rank an operational meaning, as the minimal non-Gaussian cost for engineering a state, in terms of single-photon additions. 
We have derived the equivalence classes for Gaussian convertibility using the notion of core states, and we have studied in detail the robustness of the ranks of the stellar hierarchy. Finally, we have introduced the smoothed non-Gaussianity of formation as a robust alternative to the stellar rank, in the context of approximate state engineering, and quantum computing with continuous variables.

The robustness of the genuine $n$-photon non-Gaussian hierarchy has been investigated numerically in~\cite{lachman2018faithful}.
We demonstrated analytically this robustness in this Letter, and provided an explicit method for computing the stellar robustness of any finite stellar rank state. This allows characterizing the threshold required for successfully certifying non-zero stellar ranks.
We expect that the robustness decreases with the rank. 
Thanks to the robustness of the stellar hierarchy, we have shown that the stellar rank can be experimentally witnessed by direct fidelity estimation with non-Gaussian target pure states. While the target state is pure, we emphasize that this certification method does not require the tested state itself to be pure.
Deriving other simple experimentally observable conditions would also be interesting, for example based on sampling from the Husimi $Q$-function with heterodyne detection~\cite{stenholm1992simultaneous}, given its relation with the stellar function. 
Another interesting perspective is to extend the stellar formalism to the case of multimode states. However, it is likely to be a challenging problem, as the stellar function for multimode states is a multivariate analytical function, which prevents the use of the factorisation theorem, crucial in the derivation of our results.


\textit{Acknowledgements}.\@ ---\@ We acknowledge funding from European Union's Horizon 2020 Research and Innovation Programme under Grant Agreement No. 820466 (CiViQ), and from the ANR-17-CE24-0035 VanQuTe.


\bibliographystyle{apsrev}
\bibliography{bibliography}


\widetext\newpage

\begin{center}
{\Large\textbf{Supplemental Material}}\\
\end{center}

\medskip\medskip

\noindent We prove in this Supplemental Material various technical results from the main text. We first introduce a few technical results, and then proceed to the formal proofs.


\section{Notations and preliminary results}
\label{app:technical}

\noindent In this section, we first recall notations and definitions from the main text, and then give some preliminary results that will be used in the following sections.\\

The set of continuous variable single-mode pure states is denoted $\Hi_\infty$. The trace distance between two states $\ket\phi$ and $\ket\psi$ will be noted $D_1(\phi,\psi)$. We also introduce the notation $\overline{\mathbb N}=\mathbb N\cup\{+\infty\}$.\\

Let $\ket\psi=\sum_{n\ge0}{\psi_n\ket n}\in\mathcal H_\infty$ be a normalised state. The \textit{stellar function} of the state $\ket\psi$ is noted $F_\psi^\star$ and is defined as
\be
\ba
F_\psi^\star(\alpha)&=e^{\frac12|\alpha|^2}\braket{\alpha^*|\psi}\\
&=\sum_{n\ge0}{\psi_n\frac{\alpha^n}{\sqrt{n!}}},
\ea
\ee
for all $\alpha\in\mathbb C$, where $\ket\alpha=e^{-\frac12|\alpha|^2}\sum_{n\ge0}{\frac{\alpha^n}{\sqrt{n!}}\ket n}\in\mathcal H_\infty$ is the coherent state of amplitude $\alpha$. 
The \textit{stellar rank} of a state $\ket\psi$ is noted $r^\star(\psi)$ and is defined as the number of zeros of its stellar function $F_\psi^\star$.
For all $N\in\overline{\mathbb N}$, the set of states of stellar rank equal to $N$ is denoted $R_N$.
For any state $\ket\psi\in\Hi_\infty$, we may write
\be
\ket\psi=\sum_{n\ge0}{\psi_n\ket n}=F_\psi^\star(\hat a^\dag)\ket0.
\label{app:Fadag}
\ee
\textit{Core states} are defined as the normalised states that have a polynomial stellar function.\\

A natural definition for the stellar rank of a mixture is the maximum (in $\bar{\mathbb N}$) of the ranks of the elements of the mixture. The stellar rank for a mixed state $\rho$ is thus defined as
\be
r^\star(\rho)=\inf_{p_i,\psi_i}\sup r^\star(\psi_i),
\label{rankmixed}
\ee
where the infimum is over the statistical ensembles such that $\rho=\sum_i{p_i\ket{\psi_i}\bra{\psi_i}}$, i.e.\@ all different mixtures equal to $\rho$.\\

The displacement operator of amplitude $\beta\in\mathbb C$ is given by $\hat D(\beta)=e^{\beta\hat a^\dag-\beta^*\hat a}$. Its action on the vacuum state yields the coherent state $\ket\beta$. The squeeze operator of parameter $\xi=re^{i\theta}\in\mathbb C$ is given by $\hat S(\xi)=e^{\frac12(\xi\hat a^2-\xi^*\hat a^{\dag2})}$. Its action on the vacuum state yields the squeezed state $\ket\xi$. 
All single-mode Gaussian operations may be decomposed as a squeezing operation and a displacement~\cite{Lloyd2012,ferraro2005gaussian}.
For any single-mode Gaussian state $\hat S(\xi)\hat D(\beta)\ket0$, where $\xi=re^{i\theta}$, the corresponding stellar function is~\cite{vourdas2006analytic}
\be
G_{\xi,\beta}^\star(\alpha)=(1-|a|^2)^{1/4}e^{-\frac12a\alpha^2+b\alpha+c},
\label{app:FSD}
\ee
where
\be
a:=e^{-i\theta}\tanh r,\text{ }b:=\beta\sqrt{1-|a|^2},\text{ }c:=\frac12a^*\beta^2-\frac12|\beta|^2.
\ee
The displacement and squeeze operators satisfy the following commutation rules:
\be
\ba
&\hat D(\beta)\hat a^\dag\hat D^\dag(\beta)=\hat a^\dag-\beta^*\\
&\hat S(\xi)\hat a^\dag\hat S^\dag(\xi)=c_r\hat a^\dag+s_re^{i\theta}\hat a,
\ea
\label{app:commutDS}
\ee
where $\xi=re^{i\theta}$, $c_r=\cosh r$, and $s_r=\sinh r$.\\

Finally, the creation and annihilation operators have the following stellar representations~\cite{vourdas2006analytic}:
\be
\hat a^\dag\rightarrow\alpha,\quad \hat a\rightarrow\partial_\alpha,
\label{app:castellar}
\ee
i.e.\@ the operator corresponding to $\hat a^\dag$ in the stellar representation is the multiplication by $\alpha$ and the operator in the stellar representation corresponding to $\hat a$ is the derivative with respect to $\alpha$. This means for example that the stellar function of a state $\hat a\ket\psi$ is given by $\alpha\mapsto\partial_\alpha F_\psi^\star(\alpha)$.


\section{Unicity of the stellar function}
\label{app:unique}

\noindent We prove in this section Lemma~\ref{unique} from the main text:

\medskip

\textit{Let $\ket\phi$ and $\ket\psi$ be pure normalised single-mode states such that $F_\phi^\star=F_\psi^\star$, up to a phase. Then $\ket\phi=\ket\psi$.
Moreover, let $\ket\chi=f(\hat a^\dag)\ket0$ be a single-mode normalised pure state, where $f$ is analytical. Then $f=F_\chi^\star$ up to a phase.}

\begin{proof}

With the notations of the Lemma, $F_\phi^\star(\alpha)=\sum_{n\ge0}{\phi_n\frac{\alpha^n}{\sqrt{n!}}}$ and $F_\psi^\star(\alpha)=\sum_{n\ge0}{\psi_n\frac{\alpha^n}{\sqrt{n!}}}$. The functions $F_\phi^\star$ and $F_\psi^\star$ are analytical, so $F_\phi^\star(\alpha)=F_\psi^\star(\alpha)$ implies that $\phi_n=\psi_n$ for all $n\ge0$ (up to a global phase). Hence $\ket\phi=\ket\psi$.\\

Now with $\ket\chi=\sum_{n\ge0}{\chi_n\ket n}=f(\hat a^\dag)\ket0$, let us write $f(z)=\sum_{n\ge0}{f_nz^n}$. We obtain
\be
\ba
\ket\chi&=\sum_{n\ge0}{f_n(\hat a^\dag)^n\ket0}\\
&=\sum_{n\ge0}{f_n\sqrt{n!}\ket n},
\ea
\ee
so $\chi_n=f_n\sqrt{n!}$ for all $n\ge0$, up to a global phase. On the other hand, for all $\alpha\in\mathbb C$,
\be
\ba
F_\chi^\star(\alpha)&=e^{\frac12|\alpha|^2}\braket{\alpha^*|\psi}\\
&=\sum_{n\ge0}{\chi_n\frac{\alpha^n}{\sqrt{n!}}}\\
&=\sum_{n\ge0}{f_n\alpha^n}\\
&=f(\alpha).
\ea
\ee

\end{proof}


\section{Decomposition of states of finite stellar rank}
\label{app:finitez}

\noindent We prove in this section Theorem~\ref{finitez} from the main text:

\medskip

\textit{Let $\ket\psi\in\bigcup_{N\in\mathbb N}{R_N}$ be a state of finite stellar rank. Let $\{\beta_1,\dots,\beta_{r^\star(\psi)}\}$ be the roots of $Q_\psi$, counted with multiplicity. Then,}
\be
\ket\psi=\frac1{\mathcal N}\left[\prod_{n=1}^{r^\star(\psi)}{\hat D(\beta_n)\hat a^\dag\hat D^\dag(\beta_n)}\right]\ket{G_\psi},
\ee
\textit{where $\hat D(\beta)$ is a displacement operator, $\ket{G_\psi}$ is a Gaussian state, and $\mathcal N$ is a normalisation constant. Moreover, this decomposition is unique up to reordering of the roots.}

\begin{proof}

We consider a state $\ket\psi$ of finite stellar rank $r^\star(\psi)\in\mathbb N$. 
Its stellar function is a holomorphic function over the complex plane, which satisfies, for all $\alpha\in\mathbb C$,
\be
\left|F_\psi^\star(\alpha)\right|=e^{\frac12|\alpha|^2}|\braket{\alpha^*|\psi}|\le e^{\frac12|\alpha|^2}.
\ee
Hence $F_\psi^\star$ has an order of growth less or equal to $2$, so by Hadamard-Weierstrass factorization theorem~\cite{stein2010complex},
\be
F_\psi^\star(\alpha)=\alpha^k\left[\prod_{n=1}^{r^\star(\psi)-k}{\left(1-\frac\alpha{\alpha_n^*}\right)e^{\frac\alpha{\alpha_n^*}+\frac12\left(\frac\alpha{\alpha_n^*}\right)^2}}\right]e^{G_0+G_1\alpha+G_2\alpha^2},
\label{HW}
\ee
for all $\mathbb\alpha\in C$, where $k\in\mathbb N$ is the multiplicity of $0$ as a root of $F_\psi^\star$, where the $\{\alpha_n\}$ are the non-zero roots of $Q_\psi$ counted with multiplicity (i.e.\@ the $\{\alpha_n^*\}$ are the non-zero roots of $F^\star_\psi$ counted with multiplicity), and where $G_0,G_1,G_2\in\mathbb C$. Let us introduce for brevity $M=r^\star(\psi)-k\in\mathbb N$. Because the product in the above equation is finite, we need not worry about convergence of individual factors, and we may reorder the expression at will. We obtain
\be
\ba
F_\psi^\star(\alpha)&=\alpha^k\prod_{n=1}^M{\left(1-\frac\alpha{\alpha_n^*}\right)}\cdot\prod_{n=1}^M{e^{\frac\alpha{\alpha_n^*}+\frac12\left(\frac\alpha{\alpha_n^*}\right)^2}}\cdot e^{G_0+G_1\alpha+G_2\alpha^2}\\
&=\alpha^k\prod_{n=1}^M{\left(1-\frac\alpha{\alpha_n^*}\right)}\cdot e^{G_0+\left(G_1+\sum_{n=1}^M{\frac1{\alpha_n^*}}\right)\alpha+\left(G_2+\frac12\sum_{n=1}^M{\frac1{(\alpha_n^*)^2}}\right)\alpha^2}\\
&=\frac{(-1)^M}{\prod_{n=1}^M{\alpha_n^*}}\left[\alpha^k\prod_{n=1}^M{\left(\alpha-\alpha_n^*\right)}\right]\cdot e^{G_0+\left(G_1+\sum_{n=1}^M{\frac1{\alpha_n^*}}\right)\alpha+\left(G_2+\frac12\sum_{n=1}^M{\frac1{(\alpha_n^*)^2}}\right)\alpha^2}.
\ea
\ee
With Eqs.~(\ref{app:Fadag},\ref{app:commutDS}), we obtain, for all $\beta\in\mathbb C$,
\be
\ba
\ket\psi&=F_\psi^\star(\hat a^\dag)\ket0\\
&=\frac{(-1)^M}{\prod_{n=1}^M{\alpha_n^*}}\left[(\hat a^\dag)^k\prod_{n=1}^M{\left(\hat a^\dag-\alpha_n^*\right)}\right]\cdot e^{G_0+\left(G_1+\sum_{n=1}^M{\frac1{\alpha_n^*}}\right)\hat a^\dag+\left(G_2+\frac12\sum_{n=1}^M{\frac1{(\alpha_n^*)^2}}\right)(\hat a^\dag)^2}\ket0\\
&=\frac{(-1)^M}{\prod_{n=1}^M{\alpha_n^*}}\left[(\hat a^\dag)^k\prod_{n=1}^M{\hat D(\alpha_n)\hat a^\dag\hat D^\dag(\alpha_n)}\right]\cdot e^{G_0+\left(G_1+\sum_{n=1}^M{\frac1{\alpha_n^*}}\right)\hat a^\dag+\left(G_2+\frac12\sum_{n=1}^M{\frac1{(\alpha_n^*)^2}}\right)(\hat a^\dag)^2}\ket0.
\ea
\ee
Regrouping the non-zero roots $\{\alpha_n\}$ and the $k^{th}$ zero roots into the set of zeros counted with multiplicity $\{\beta_n\}$, we obtain
\be
\ket\psi=\frac{(-1)^M}{\prod_{n=1}^M{\alpha_n^*}}\left[\prod_{n=1}^{r^\star(\psi)}{\hat D(\beta_n)\hat a^\dag\hat D^\dag(\beta_n)}\right]\cdot e^{G_0+\left(G_1+\sum_{n=1}^M{\frac1{\alpha_n^*}}\right)\hat a^\dag+\left(G_2+\frac12\sum_{n=1}^M{\frac1{(\alpha_n^*)^2}}\right)(\hat a^\dag)^2}\ket0.
\ee
Finally, the state
\be
e^{G_0+\left(G_1+\sum_{n=1}^M{\frac1{\alpha_n^*}}\right)\hat a^\dag+\left(G_2+\frac12\sum_{n=1}^M{\frac1{(\alpha_n^*)^2}}\right)(\hat a^\dag)^2}\ket0
\ee
is a (non normalised) Gaussian state, by Eq.~(\ref{app:FSD}) and Lemma~\ref{unique}. We finally obtain
\be
\ket\psi=\frac1{\mathcal N}\left[\prod_{n=1}^{r^\star(\psi)}{\hat D(\beta_n)\hat a^\dag\hat D^\dag(\beta_n)}\right]\ket{G_\psi},
\ee
where $\mathcal N$ is a normalisation constant, and $\ket{G_\psi}$ is a Gaussian state. The decomposition is unique by Lemma~\ref{unique} (up to a global phase and a reordering of the roots).

\end{proof}


\section{Operations leaving the stellar rank invariant}
\label{app:Ginv}

\noindent We prove in this section Theorem~\ref{Ginv} from the main text:

\medskip

\textit{A unitary operation is Gaussian if and only if it leaves the stellar rank invariant.}

\begin{proof}

If a unitary operation leaves the stellar rank invariant, it maps in particular all pure states of stellar rank zero to pure states of stellar rank zero, i.e.\@ all Gaussian states to Gaussian states, so it is a Gaussian operation.\\

Reciprocally, let us show that Gaussian unitary operations leave the stellar rank invariant.
We first consider finite stellar rank pure states. Let $\ket \psi$ be such a state. By Theorem~\ref{finitez},
\be
\ket\psi=P_\psi(\hat a^\dag)\ket{G_\psi},
\ee
where $P_\psi$ is a polynomial of degree $r^\star(\psi)$ and $\ket{G_\psi}$ is a Gaussian state. By Eq.~(\ref{app:commutDS}) and by linearity we have
\be
\ket{\psi_\beta}:=\hat D(\beta)\ket\psi=\hat P_\psi(\hat a^\dag-\beta^*)\hat D(\beta)\ket{G_\psi},
\ee
and
\be
\ket{\psi_\xi}:=\hat S(\xi)\ket\psi=P_\psi(c_r\hat a^\dag+s_re^{i\theta}\hat a)\hat S(\xi)\ket{G_\psi},
\ee
where $\xi=re^{i\theta}$.
By Eq.~(\ref{app:castellar}), the stellar operator corresponding to $\hat a^\dag$ is the multiplication by $\alpha$, and the stellar operator corresponding to $\hat a$ is the derivative with respect to $\alpha$. In particular, during a displacement of $\beta$, the stellar function of $\ket\psi$ is modified as
\be
F_\psi^\star(\alpha)\rightarrow F_{\psi_\beta}^\star(\alpha)=e^{\alpha\beta-\frac12|\beta|^2}F_\psi^\star(\alpha-\beta^*)=P_\psi(\alpha-\beta^*)G_\beta^\star(\alpha),
\label{Fdisplaced}
\ee
where $G_\beta^\star(\alpha)$ is the Gaussian stellar function corresponding to the Gaussian state $\hat D(\beta)\ket{G_\psi}$. During a squeezing of $\xi$, the stellar function of $\ket\psi$ is modified as
\be
\ba
F_\psi^\star(\alpha)\rightarrow F_{\psi_\xi}^\star(\alpha)&=P_\psi\left(c_r\alpha+s_re^{i\theta}\partial_\alpha\right)G_\xi^\star(\alpha)\\
&=Q_{\psi,r}(\alpha)G_\xi^\star(\alpha),
\ea
\ee
where $G_\xi^\star(\alpha)$ is the Gaussian stellar function corresponding to the Gaussian state $\hat S(\xi)\ket{G_\psi}$, and where $Q_{\psi,\xi}(\alpha)=G_\xi^{\star-1}(\alpha)\left[P_\psi\left(c_r\alpha+s_re^{i\theta}\partial_\alpha\right)G_\xi^\star(\alpha)\right]$ is a polynomial. Let us compute the leading coefficient of $Q_{\psi,\xi}$. Writing $p$ the leading coefficient of $P_\psi$, and $N=r^\star(\psi)$ its degree for brevity, the leading coefficient of $Q_{\psi,\xi}$ is given by the leading coefficient of
\be
G_\xi^{\star-1}(\alpha)\left[p\left(c_r\alpha+s_re^{i\theta}\partial_\alpha\right)^NG_\xi^\star(\alpha)\right].
\ee
Let us write $G_\xi^\star(\alpha)=e^{-\frac12a\alpha^2+b\alpha+c}$, as in Eq.~(\ref{app:FSD}). The leading coefficient of $Q_{\psi,\xi}$ may then be obtained as the leading coefficient of
\be
e^{\frac12a\alpha^2}\left[p\left(c_r\alpha+s_re^{i\theta}\partial_\alpha\right)^Ne^{-\frac12a\alpha^2}\right].
\label{lead1}
\ee
For all $x,\lambda$, we have~\cite{wyss2017two}
\be
\left(x+\lambda\partial_x\right)^N=\sum_{n=0}^{\left\lfloor{\frac N2}\right\rfloor}{\frac{N!\lambda^n}{(N-2n)!n!2^n}\sum_{k=0}^{N-2n}{\binom{N-2n}{k}x^k\partial_x^{N-2n-k}}},
\ee
so the leading coefficient of $Q_{\psi,\xi}$ is equal to the leading coefficient of:
\be
pc_r^N\sum_{n=0}^{\left\lfloor{\frac N2}\right\rfloor}{\alpha^{N-2n}(1-a)^{N-2n}\frac{N!t_r^ne^{in\theta}}{(N-2n)!n!2^n}},
\ee
where $t_r=\tanh r$. Finally, taking the leading coefficient in $\alpha$ of this expression, corresponding to $n=0$, gives
\be
pc_r^N(1-a)^N,
\ee
which is non-zero unless $a=1$, which corresponds to an infinite value for the modulus of the squeezing parameter $\xi=re^{i\theta}$ by Eq.~(\ref{app:FSD}).
Hence the polynomials $P_\psi$ and $Q_{\psi,\xi}$ have the same degree.
This shows that a finite number of zeros is not modified by Gaussian operations.

Gaussian operations also map states with infinite number of zeros to states with infinite number of zeros. Indeed, assuming there exist a state $\ket\phi$ with an infinite number of zeros which is mapped by a Gaussian operation $\hat G$ to a state $\ket\psi$ with a finite number of zeros, then $\hat G^\dag$ would map $\ket\psi$ to $\ket\phi$, thus changing the (finite) number of zeros of $F_\psi^\star$, which would be in contradiction with the previous proof.
Hence Gaussian unitary operations leave the stellar rank of pure states invariant.

Now by Eq.~(\ref{rankmixed}), the stellar rank of a mixed state $\rho$ is given by
\be
r^\star(\rho)=\inf_{p_i,\psi_i}\sup r^\star(\psi_i),
\ee
where the infimum is over the statistical ensembles such that $\rho=\sum_i{p_i\ket{\psi_i}\bra{\psi_i}}$. For $\hat G$ a unitary Gaussian operation,
\be
\ba
r^\star(\hat G\rho\hat G^\dag)&=\inf_{p_i,\psi_i}\sup r^\star(\hat G\psi_i)\\
&=\inf_{p_i,\psi_i}\sup r^\star(\psi_i)\\
&=r^\star(\rho),
\ea
\ee
where we used in the second line the fact that Gaussian unitary operations leave the stellar rank of pure states invariant. Hence, Gaussian unitary operations leave the stellar rank invariant.

\end{proof}


\section{Gaussian convertibility}
\label{app:Gconvert}

\subsection{Proof of Theorem~\ref{Gconv}}

\noindent We prove in this section Theorem~\ref{Gconv} from the main text:

\medskip

\textit{Let $\ket\psi\in\bigcup_{N\in\mathbb N}{R_N}$ be a state of finite stellar rank. Then, there exists a unique core state $\ket{C_\psi}$ such that $\ket\psi$ and $\ket{C_\psi}$ are Gaussian-convertible.}

\textit{By Theorem~\ref{finitez}, $\ket\psi=P_\psi(\hat a^\dag)\ket{G_\psi}$, where $P_\psi$ is a polynomial of degree $r^\star(\psi)$ and $\ket{G_\psi}=\hat S(\xi)\hat D(\beta)\ket0$ is a Gaussian state, where $\hat D(\beta)=e^{\beta\hat a^\dag-\beta^*\hat a}$ is a displacement operator, and $\hat S(\xi)=e^{\frac12(\xi\hat a^2-\xi^*\hat a^{\dag2})}$ is a squeezing operator, with $\xi=re^{i\theta}$. Then,}
\be
\ket\psi=\hat S(\xi)\hat D(\beta)\ket{C_\psi}=\hat S(\xi)\hat D(\beta)F_{C_\psi}^\star(\hat a^\dag)\ket0,
\ee
\textit{where the (polynomial) stellar function of $\ket{C_\psi}$ is given by}
\be
F_{C_\psi}^\star(\alpha)=P_\psi\left(c_r\alpha-s_re^{i\theta}\partial_\alpha+c_r\beta^*-s_re^{i\theta}\beta\right)\cdot1,
\ee
\textit{for all $\alpha\in\mathbb C$.}

\begin{proof}

Let $\ket\psi\in\bigcup_{N\in\mathbb N}{R_N}$ be a state of finite stellar rank. By Theorem~\ref{finitez},
\be
\ket\psi=P_\psi(\hat a^\dag)\ket{G_\psi},
\label{decomppsi}
\ee
where $P_\psi$ is a polynomial of degree $r^\star(\psi)$ and $\ket{G_\psi}=\hat S(\xi)\hat D(\beta)\ket0$ is a Gaussian state, with $\xi=re^{i\theta}$. Let us define $\ket{C_\psi}=\hat D^\dag(\beta)\hat S^\dag(\xi)\ket\psi$. The states $\ket\psi$ and $\ket{C_\psi}$ are Gaussian-convertible. Moreover, from the commutation relations in Eq.~(\ref{app:commutDS}) and by linearity we obtain
\be
\ba
\ket{C_\psi}&=\hat D^\dag(\beta)\hat S^\dag(\xi)P_\psi(\hat a^\dag)\ket{G_\psi}\\
&=\hat D^\dag(\beta)P_\psi\left(c_r\hat a^\dag-s_re^{i\theta}\hat a\right)\hat S^\dag(\xi)\ket{G_\psi}\\
&=P_\psi\left[c_r(\hat a^\dag+\beta^*)-s_re^{i\theta}(\hat a+\beta)\right]\ket0\\
&=P_\psi\left(c_r\hat a^\dag-s_re^{i\theta}\hat a+c_r\beta^*-s_re^{i\theta}\beta\right)\ket0,
\ea
\ee
where we used Eq.~(\ref{decomppsi}) in the first line. By Eq.~(\ref{app:castellar}), the stellar operator corresponding to $\hat a^\dag$ is the multiplication by $\alpha$ and the stellar operator corresponding to $\hat a$ is the derivative with respect to $\alpha$. Hence,
\be
F_{C_\psi}^\star(\alpha)=P_\psi\left(c_r\alpha-s_re^{i\theta}\partial_\alpha+c_r\beta^*-s_re^{i\theta}\beta\right)\cdot1,
\ee
for all $\alpha\in\mathbb C$, which is a polynomial function, so the state $\ket{C_\psi}$ is a core state.

In order to conclude the proof, we need to show that $\ket{C_\psi}$ is the unique core state Gaussian-convertible to $\ket\psi$. Let $\ket C=P_C(\hat a^\dag)\ket0$ be a core state Gaussian-convertible to $\ket\psi$. The states $\ket{C_\psi}$ and $\ket C$ are Gaussian-convertible so there exist $\xi,\beta\in\mathbb C$ such that
\be
\ba
\ket{C_\psi}&=\hat S(\xi)\hat D(\beta)\ket C\\
&=\hat S(\xi)\hat D(\beta)P_C(\hat a^\dag)\ket0\\
&=P_C(c_r\hat a^\dag+s_re^{i\theta}\hat a-\beta^*)\hat S(\xi)\hat D(\beta)\ket0,
\ea
\ee
where we used Eq.~(\ref{app:commutDS}). Hence,
\be
F_{C_\psi}^\star(\alpha)=P_C(c_r\alpha+s_re^{i\theta}\partial_\alpha-\beta^*)G_{\xi,\beta}^\star(\alpha).
\ee
With Eq.~(\ref{app:FSD}), this function may be expressed as a polynomial multiplied by a Gaussian function $G_{\xi,\beta}^\star$. On the other hand $F_{C_\psi}^\star$ is a polynomial, since $\ket{C_\psi}$ is a core state. By comparison of the speed of convergence, this implies that the Gaussian function $G_{\xi,\beta}^\star$ is constant, i.e.\@ that
\be
e^{-i\theta}\tanh r=0\text{ and }\beta\sqrt{1-\tanh^2r}=0,
\ee
by Eq.~(\ref{app:FSD}). This in turn implies $\xi=\beta=0$, and $\ket C=\hat S(\xi)\hat D(\beta)\ket C=\ket{C_\psi}$.

\end{proof}


\subsection{Gaussian convertibility, a simple example}
\label{app:exGconv}

\noindent We consider the following simple example to illustrate the use of Theorem~\ref{Gconv} for determining Gaussian convertibility: a photon-subtracted squeezed state and a single-photon Fock state.
We write $\ket\phi=-\frac1{s_\xi}\hat a\ket{\xi}$ a normalised photon-subtracted squeezed vacuum state, with $\xi\in\mathbb R$ and $s_\xi=\sinh\xi$. We write also $\ket\psi=\ket1$ a single-photon Fock state. Using Eq.~(\ref{app:FSD}) and Eq.~(\ref{app:castellar}), we obtain for all $\alpha\in\mathbb C$
\be
\ba
F_\phi^\star(\alpha)&=-\frac1{s_\xi}\partial_\alpha\left[e^{-\frac12t_\xi\alpha^2}\right]\\
&=\frac\alpha{c_\xi}e^{-\frac12t_\xi\alpha^2},
\ea
\ee
where $c_\xi=\cosh\xi$ and $t_\xi=\tanh\xi$. We also have $F_\psi^\star(\alpha)=\alpha$. With the notations of Theorem~\ref{Gconv}, we have $r_\phi=\xi$, $r_\psi=\theta_\phi=\theta_\psi=\beta_\phi=\beta_\psi=0$, $\hat G_\phi=\hat S(\xi)$, $\hat G_\psi=\hat{\mathds 1}$, $P_\phi(\alpha)=\frac\alpha{c_\xi}$, and $P_\psi(\alpha)=\alpha$, so for all $\alpha\in\mathbb C$,
\be
\ba
P_\phi\left(c_{r_\phi}\alpha-s_{r_\phi}e^{i\theta_\phi}\partial_\alpha+c_{r_\phi}\beta_\phi^*-s_{r_\phi}e^{i\theta_\phi}\beta_\phi\right)\cdot1&=\frac1{c_\xi}\left(c_\xi\alpha-s_\xi\partial_\alpha\right)\cdot1\\
&=\alpha,
\ea
\ee
and
\be
\ba
P_\psi\left(c_{r_\psi}\alpha-s_{r_\psi}e^{i\theta_\psi}\partial_\alpha+c_{r_\psi}\beta_\psi^*-s_{r_\psi}e^{i\theta_\psi}\beta_\psi\right)\cdot1&=\alpha\cdot1\\
&=\alpha,
\ea
\ee
thus $\ket\phi$ and $\ket\psi$ share the same core state. By Theorem~\ref{Gconv}, this means that $\ket\phi$ and $\ket\psi$ are Gaussian-convertible, and we have $\ket\phi=\hat G_\phi\hat G_\psi^\dag\ket\psi$, where
\be
\hat G_\phi\hat G_\psi^\dag=\hat S(\xi).
\ee
Using Eq.~(\ref{app:commutDS}) confirms indeed that $-\frac1{s_\xi}\hat a\ket{\xi}=\hat S(\xi)\ket1$.


\section{Stellar robustness}


\subsection{Proof of Lemma~2}

\noindent We prove in this section Lemma~2 from the main text:

\medskip

\textit{Let $\ket\psi\in\mathcal H_\infty$, then}
\begin{equation}
\sup_{r^\star(\rho)<r^\star(\psi)}F(\rho,\psi)=1-[R^\star(\psi)]^2,
\end{equation}
\textit{where $F$ is the fidelity.}

\begin{proof}
For any pure state $\ket\psi\in\Hi_\infty$, and any set of pure states $\mathcal X$, we have
\be
\ba
\sup_{\substack{\rho=\sum{p_i{\ket\phi_i}\bra{\phi_i}}\\ \sum{p_i}=1,\phi_i\in\mathcal X}}{F(\rho,\psi)}&=\sup_{\substack{\rho=\sum{p_i{\ket\phi_i}\bra{\phi_i}}\\ \sum{p_i}=1,\phi_i\in\mathcal X}}{\braket{\psi|\rho|\psi}}\\
&=\sup_{\sum{p_i}=1}\sup_{\phi_i\in\mathcal X}{\sum{p_i|\braket{\phi_i|\psi}|^2}}\\
&=\sup_{\phi\in\mathcal X}{|\braket{\phi|\psi}|^2}\\
&=\sup_{\phi\in\mathcal X}{F(\phi,\psi)}.
\ea
\ee
Hence, for $\mathcal X$ the set of pure states of stellar rank strictly smaller than $r^\star(\psi)$,
\be
\ba
R^\star(\psi)&=\inf_{r^\star(\phi)<r^\star(\psi)}{D_1(\phi,\psi)}\\
&=\inf_{r^\star(\phi)<r^\star(\psi)}{\sqrt{1-|\braket{\phi|\psi}|^2}}\\
&=\sqrt{1-\sup_{r^\star(\phi)<r^\star(\psi)}{F(\phi,\psi)}}\\
&=\sqrt{1-\sup_{r^\star(\rho)<r^\star(\psi)}{F(\rho,\psi)}},
\ea
\ee
where $D_1$ denotes the trace distance, where we used the definition of the stellar rank for mixed states~(\ref{rankmixed}). We finally obtain
\be
\sup_{r^\star(\rho)<r^\star(\psi)}{F(\rho,\psi)}=1-[R^\star(\psi)]^2.
\ee

\end{proof}


\subsection{Stellar robustness of single photon-added squeezed states}

\noindent We consider a (normalised) single photon-added squeezed state $\frac1{c_r}\hat a^\dag\ket\xi$, for $\xi=re^{i\theta}\in\mathbb C$, where $c_r=\cosh r$. By Eq.~(\ref{app:commutDS}), we have
\be
\frac1{c_r}\hat a^\dag\ket\xi=\hat S(\xi)\ket1,
\ee
and since the stellar robustness is invariant under Gaussian operations, the stellar robustness of a single photon-added squeezed state is equal to the stellar robustness of a single-photon Fock state (and in particular is independent of the squeezing parameter). In the following, we compute analytically this robustness.

The single-photon Fock state $\ket1$ is a core state of stellar rank $1$. Its stellar robustness is thus given by the minimal distance with the set of rank $0$ states, i.e.\@ the set $\mathcal G$ of Gaussian states. Since any single-mode Gaussian state may be obtained from the vacuum by a displacement and a squeezing, we obtain
\be
\ba
R^\star(\ket1)&=\inf_{\ket\phi\in\mathcal G}{D_1(\ket\phi,\ket1)}\\
&=\inf_{\ket\phi\in\mathcal G}{\sqrt{1-|\braket{\phi|1}|^2}}\\
&=\inf_{\xi,\gamma\in\mathbb C}{\sqrt{1-|\braket{0|\hat D(\gamma)\hat S(\xi)|1}|^2}}\\
&=\sqrt{1-\sup_{\xi,\gamma\in\mathbb C}{|\braket{0|\hat D(\gamma)\hat S(\xi)|1}|^2}}.
\ea
\ee
With Eq.~(\ref{app:commutDS}) we obtain $\hat D(\gamma)\hat S(\xi)\ket1=\frac1{c_r}(\hat a^\dag-\gamma^*)\hat D(\gamma)\hat S(\xi)\ket0$, hence
\be
\ba
R^\star(\ket1)&=\sqrt{1-\sup_{\xi,\gamma\in\mathbb C}{\left[\frac1{c_r^2}|\braket{0|(\hat a^\dag-\gamma^*)\hat D(\gamma)\hat S(\xi)|0}|^2\right]}}\\
&=\sqrt{1-\sup_{\xi,\gamma\in\mathbb C}{\left[\frac{|\gamma|^2}{c_r^2}|\braket{0|\hat D(\gamma)\hat S(\xi)|0}|^2\right]}}\\
&=\sqrt{1-\sup_{\xi,\gamma\in\mathbb C}{\left[\frac{|\gamma|^2}{c_r^2}|\braket{-\gamma|\xi}|^2\right]}}.
\ea
\ee
where $\ket{-\gamma}$ is a coherent state and $\ket\xi$ a squeezed state. Now for all $\gamma,\xi=re^{i\theta}\in\mathbb C$, we have
\be
|\braket{-\gamma|\xi}|^2=\frac1{c_r}e^{-|\gamma|^2-\frac{t_r}2(\gamma^2e^{i\theta}+\gamma^{*2}e^{-i\theta})}.
\ee
where $c_r=\cosh r$ and $t_r=\tanh r$. Setting $\tilde\gamma=\gamma e^{i\theta/2}$ gives
\be
|\braket{-\gamma|\xi}|^2=\frac1{c_r}e^{-|\tilde\gamma|^2-\frac{t_r}2(\tilde\gamma^2+\tilde\gamma^{*2})},
\ee
so that
\be
R^\star(\ket1)=\sqrt{1-\sup_{r\ge0,\tilde\gamma\in\mathbb C}{\left[\frac{|\tilde\gamma|^2}{c_r^3}e^{-|\tilde\gamma|^2-\frac{t_r}2(\tilde\gamma^2+\tilde\gamma^{*2})}\right]}}.
\ee
Writing $\tilde\gamma=x+iy$ we obtain
\be
\ba
\frac{|\tilde\gamma|^2}{c_r^3}e^{-|\tilde\gamma|^2-\frac{t_r}2(\tilde\gamma^2+\tilde\gamma^{*2})}&=\frac{x^2+y^2}{c_r^3}e^{-(1+t_r)x^2}e^{-(1-t_r)y^2}\\
&=(1-t_r^2)^{3/2}(x^2+y^2)e^{-(1+t_r)x^2}e^{-(1-t_r)y^2}\\
&=(1-t_r^2)^{3/2}(x^2+y^2)e^{-(1-t_r)(x^2+y^2)}e^{-2t_rx^2}\\
&\le(1-t_r^2)^{3/2}(x^2+y^2)e^{-(1-t_r)(x^2+y^2)}\\
&\le\frac{(1-t_r^2)^{3/2}}{e(1-t_r)}\\
&=\frac1e\sqrt{(1-t_r)(1+t_r)^3},
\ea
\ee
and this bound is attained for $x=0$ and $y=\frac1{\sqrt{1-t_r}}$.
Finally, we have $\max_{u\in[0,1]}{(1-u)(1+u)^3}=\frac{27}{16}$, attained for $u=\frac12$, so we deduce the stellar robustness of a single photon-added squeezed state:
\be
R^\star(\hat a^\dag\ket\xi)=R^\star(\ket1)=\sqrt{1-\frac{3\sqrt3}{4e}}\approx0.72.
\ee
%


\subsection{Stellar robustness of finite-rank states}

\noindent In this section we reduce the computation of the stellar robustness of finite stellar rank states to an optimization problem.

Let $\ket\psi\in\Hi_\infty$ be a non-Gaussian pure state of finite stellar rank $r^\star(\psi)\in\mathbb N^*$. For $N\in\mathbb N$, let us define
\be
\tilde{\mathbb C}_n[X]:=\left\{P\in\mathbb C[X]\text{ s.t. }P(X)=\sum_{k=0}^n{\frac{p_k}{\sqrt{k!}}X^k}\text{ and }\sum_{k=0}^n{|p_k|^2}=1\right\}.
\ee

By Theorem~\ref{Gconv}, there exist a unique polynomial $P_\psi\in\tilde{\mathbb C}_{r^\star(\psi)}[X]$ and a unique Gaussian operation $\hat G_\psi$ such that $\ket\psi=\hat G_\psi P_\psi(\hat a^\dag)\ket0$.
Similarly, for any pure state $\ket\phi$ such that $r^\star(\phi)<r^\star(\psi)$, there exist a unique polynomial $P_\phi\in\tilde{\mathbb C}_{r^\star(\phi)}[X]$ and a unique Gaussian operation $\hat G_\phi$ such that $\ket\phi=\hat G_\phi P_\phi(\hat a^\dag)\ket0$.
The stellar robustness of the state $\ket\psi$ is then given by
\be
\ba
R^\star(\psi)&=\inf_{r^\star(\phi)<r^\star(\psi)}{D_1(\phi,\psi)}\\
&=\sqrt{1-\sup_{r^\star(\phi)<r^\star(\psi)}{|\braket{\psi|\phi}|^2}}\\
&=\sqrt{1-\sup_{r^\star(\phi)<r^\star(\psi)}{|\braket{0|P_\psi^*(\hat a)\hat G_\psi^\dag\hat G_\phi P_\phi(\hat a^\dag)|0}|^2}}\\
&=\sqrt{1-\sup_{\substack{\xi,\beta\in\mathbb C\\Q\in\tilde{\mathbb C}_{r^\star(\psi)-1}[X]}}{|\braket{0|P_\psi^*(\hat a)\hat S(\xi)\hat D(\beta)Q(\hat a^\dag)|0}|^2}}.
\ea
\ee
For $\xi=re^{i\theta},\beta\in\mathbb C$, and $Q\in\mathbb C[X]$, we have
\be
\hat S(\xi)\hat D(\beta)Q(\hat a^\dag)\ket0=Q(\cosh(r)\hat a^\dag+\sinh(r)e^{i\theta}\hat a-\beta^*)\hat S(\xi)\hat D(\beta)\ket0,
\label{stateinter}
\ee
and by Eq.~(\ref{app:FSD}), the stellar function of the Gaussian state $\hat S(\xi)\hat D(\beta)\ket0$ is given by
\be
G_{\xi,\beta}^\star(\alpha)=(1-|a|^2)^{1/4}e^{-\frac12a\alpha^2+b\alpha+c},
\ee
where
\be
a=e^{-i\theta}\tanh r,\text{ }b=\beta\sqrt{1-|a|^2},\text{ }c=\frac12a^*\beta^2-\frac12|\beta|^2.
\ee
Hence, the stellar function of the state in Eq.~(\ref{stateinter}) is given by
\be
F(\alpha)=(1-|a|^2)^{1/4}Q(\cosh(r)\alpha+\sinh(r)e^{i\theta}\partial_\alpha-\beta^*)e^{-\frac12a\alpha^2+b\alpha+c}.
\ee
For all state $\ket\chi$, we have $F_\chi(0)=\braket{0|\chi}$. We thus finally get
\be
R^\star(\psi)=\sqrt{1-\sup_{\substack{\xi,\beta\in\mathbb C\\Q\in\tilde{\mathbb C}_{r^\star(\psi)-1}[X]}}{\sqrt{1-|a|^2}\left|\left[P_\psi^*(\partial_\alpha)Q(\cosh(r)\alpha+\sinh(r)e^{i\theta}\partial_\alpha-\beta^*)e^{-\frac12a\alpha^2+b\alpha+c}\right]_{\alpha=0}\right|^2}},
\ee
where
\be
\xi=re^{i\theta},\text{ }a=e^{-i\theta}\tanh r,\text{ }b=\beta\sqrt{1-|a|^2},\text{ }c=\frac12a^*\beta^2-\frac12|\beta|^2.
\ee
Plugging these expressions in the above equation, and writing $c_r=\cosh r$, $s_r=\sinh r$, and $t_r=\tanh r$ for brevity, we finally get
\be
R^\star(\psi)=\sqrt{1-\sup_{\substack{\xi=re^{i\theta},\beta\in\mathbb C\\Q\in\tilde{\mathbb C}_{r^\star(\psi)-1}[X]}}{\frac1{c_r}\left|\left[P_\psi^*(\partial_\alpha)Q(c_r\alpha+s_re^{i\theta}\partial_\alpha-\beta^*)e^{-\frac{t_r}2e^{-i\theta}\alpha^2+\frac1{c_r}\alpha+\frac{t_r}2e^{i\theta}\beta^2-\frac12|\beta|^2}\right]_{\alpha=0}\right|^2}},
\ee
where we used $1-t_r^2=\frac1{c_r^2}$. The supremum in this last expression may be obtained analytically for low stellar rank states, as it is done in the previous section, and numerically in the general case.


\section{Topology of the stellar hierarchy}
\label{app:topology}

\noindent We prove in this section Theorem~\ref{topology} from the main text:

\medskip

\textit{For all $N\in\mathbb N$,}
\be
\overline{R_N}=\underset{0\le K\le N}{\bigcup}{R_K},
\ee
\textit{where $\overline X$ denotes the closure of $X$ for the trace norm in the set of normalised states $\mathcal H_\infty$.}

\begin{proof}

Recall that the set of normalised pure single-mode states is closed for the trace norm in the whole Hilbert space, since it is the reciprocal image of $\{1\}$ by the trace norm, which is Lipschitz continuous with Lipschitz constant $1$, hence continuous.\\

For the proof, we fix $N\in\mathbb N$. We prove the theorem by showing a double inclusion. We first show that $\bigcup_{K=0}^N{R_K}\subset\overline{R_N}$,
and then that the set $\bigcup_{K=0}^N{R_K}$ is closed in $\mathcal H_\infty$ for the trace norm. Since the closure of a set $X$ is the smallest closed set containing $X$, and given that $R_N\subset\bigcup_{K=0}^N{R_K}$, this will prove the other inclusion and hence the result.\\

We have $R_N\subset\overline{R_N}$. Let $\ket\psi\in\bigcup_{K=0}^{N-1}{R_K}$. There exists $K\in\{0,\dots,N-1\}$ such that $r^\star(\psi)=K$. By Theorem~\ref{Gconv}, there exists a core state $\ket{C_\psi}$, with a polynomial stellar function of degree $K$, and a Gaussian operation $\hat G_\psi$ such that $\ket\psi=\hat G_\psi\ket{C_\psi}$. We define the sequence of normalised states
\be
\ket{\psi_m}=\sqrt{1-\frac1m}\ket\psi+\frac1{\sqrt m}\hat G_\psi\ket N,
\ee
for $m\ge1$. We have
\be
\ket{\psi_m}=\hat G_\psi\left(\sqrt{1-\frac1m}\ket{C_\psi}+\frac1{\sqrt m}\ket N\right),
\ee
and the state $\sqrt{1-\frac1m}\ket{C_\psi}+\frac1{\sqrt m}\ket N$ is a normalised core state whose stellar function is a polynomial of degree $N$, hence $\ket{\psi_m}\in R_N$. Moreover, $\{\ket{\psi_m}\}_{m\ge1}$ converges to $\ket\psi$ in trace norm. This shows that $\bigcup_{K=0}^N{R_K}\subset\overline{R_N}$.\\

We now prove that the set $\bigcup_{K=0}^N{R_K}$ is closed in $\mathcal H_\infty$ for the trace norm. For $N=0$ (i.e.\@ the set of Gaussian states is a closed set), this is already a non-trivial result, and a proof may be found e.g.\@ in~\cite{lami2018gaussian}. 

For all $N\ge0$, the proof is a bit more involved. 
The sketch of the proof is the following: given a converging sequence in $\bigcup_{K=0}^N{R_K}$, we want to show that its limit has a stellar rank less or equal to $N$. We first use the decomposition result of Theorem~\ref{Gconv}, in order to obtain a sequence of Gaussian operations acting on a sequence of core states of rank less or equal to $N$. We make use of the compactness of this set of core states to restrict to a unique core state. Then, we show that the squeezing and the displacement parameters of the sequence of Gaussian operations cannot be unbounded. This allows to conclude by extracting converging subsequences from these parameters.\\

Let us write $D$ the trace distance, induced by the trace norm. Let $\{\ket{\psi_m}\}_{m\in\mathbb N}\in\bigcup_{K=0}^N{R_K}$ be a converging sequence for the trace norm, and let $\ket\psi\in\mathcal H_\infty$ be its limit. By Theorem~\ref{Gconv}, there exist a sequence of core states $\{\ket{C_m}\}_{m\in\mathbb N}$, with polynomial stellar functions of degrees less or equal to $N$, and a sequence of Gaussian operations $\{\hat G_m\}_{m\in\mathbb N}$ such that for all $m\in\mathbb N$, $\ket{\psi_m}=\hat G_m\ket{C_m}$.

The set of normalised core states with a polynomial stellar function of degree less or equal to $N$ corresponds to the set of normalised states with a support over the Fock basis truncated at $N$, and is compact for the trace norm in $\mathcal H_\infty$ (isomorphic to the norm $1$ vectors in $\mathbb C^{N+1}$). Hence, the sequence $\{\ket{C_m}\}_{m\in\mathbb N}$ admits a converging subsequence $\{\ket{C_{m_k}}\}_{k\in\mathbb N}$. Let the core state $\ket C$, with a polynomial stellar function of degree less or equal to $N$, be its limit. Along this subsequence,
\be
\ket{\psi_{m_k}}=\hat G_{m_k}\ket{C_{m_k}},
\label{psimk}
\ee
and we have $\text{lim}_{k\rightarrow+\infty}D(\ket{\psi_{m_k}},\ket\psi)=0$ and $\text{lim}_{k\rightarrow+\infty}D(\ket{C_{m_k}},\ket C)=0$. Moreover, for all $k\in\mathbb N$,
\be
\ba
D(\hat G_{m_k}\ket C,\ket\psi)&\le D(\hat G_{m_k}\ket C,\ket{\psi_{m_k}})+D(\ket{\psi_{m_k}},\ket\psi)\\
&=D(\hat G_{m_k}\ket C,\hat G_{m_k}\ket{C_{m_k}})+D(\ket{\psi_{m_k}},\ket\psi)\\
&=D(\ket C,\ket{C_{m_k}})+D(\ket{\psi_{m_k}},\ket\psi),
\ea
\ee
where we used the triangular inequality in the first line, Eq.~(\ref{psimk}) in the second line, and the invariance of the trace distance under unitary transformations in the third line. Hence, the sequence $\{\hat G_{m_k}\ket C\}_{k\in\mathbb N}$ converges in trace norm to $\ket\psi$. This shows that we can restrict without loss of generality to a unique core state, with a polynomial stellar function of degree less or equal to $N$, instead of a sequence of such core states.\\

Let $\ket C$ thus be a core state, with a polynomial stellar function of degree $K$ less or equal to $N$. We write
\be
\ket C=P_C(\hat a^\dag)\ket0=\sum_{n=0}^K{\frac{p_n}{\sqrt{n!}}\ket n},
\label{coreC}
\ee
with $\sum_{n=0}^K{\frac{|p_n|^2}{n!}}=1$. Let us consider a converging sequence $\{\hat G_m\ket C\}_{m\in\mathbb N}$, where $\hat G_m$ are Gaussian operations, and denote $\ket\psi$ its limit. There exists two sequences $\{\xi_m\}_{m\in\mathbb N}$ and $\{\beta_m\}_{m\in\mathbb N}$, such that for all $m\in\mathbb N$,
\be
\hat G_m=\hat S(\xi_m)\hat D(\beta_m).
\ee
We write $\xi_m=r_me^{i\theta_m}$, with $r_m\ge0$, for all $m\in\mathbb C$. We may rewrite $\hat G_m=\hat D(\gamma_m)\hat S(\xi_m)$, where for all $m\in\mathbb N$,
\be
\gamma_m=c_{r_m}\beta_m+s_{r_m}e^{i\theta_m}\beta_m^*,
\ee
where $c_{r_m}=\cosh(r_m)$ and $s_{r_m}=\sinh(r_m)$. With these notations, we prove the following result:\\

\textit{Lemma.}
The sequences $\{\xi_m\}_{m\in\mathbb N}$ and $\{\gamma_m\}_{m\in\mathbb N}$ are bounded.\\

\textit{Proof}.---We start by computing a useful upper bound for the $Q$-function of the state $\hat G_m\ket C$, which we obtain in Eq.~(\ref{boundQ3}). For $m\in\mathbb N$, we have:
\be
\ba
Q_{\hat G_m\ket C}(\alpha)&=Q_{\hat D(\gamma_m)\hat S(\xi_m)\ket C}(\alpha)\\
&=Q_{\hat S(\xi_m)\ket C}(\alpha-\gamma_m)\\
&=\frac{e^{-|\alpha-\gamma_m|^2}}\pi\left|F_{\hat S(\xi_m)\ket C}^\star(\alpha^*-\gamma_m^*)\right|^2,
\ea
\label{QGmC1}
\ee
for all $\alpha\in\mathbb C$. We have
\be
\ba
\hat S(\xi_m)\ket C&=\hat S(\xi_m)P_C(\hat a^\dag)\ket0\\
&=P_C(c_{r_m}\hat a^\dag+s_{r_m}e^{i\theta_m}\hat a)\hat S(\xi_m)\ket0.
\ea
\ee
Hence, with Eq.~(\ref{app:FSD},\ref{app:castellar}),
\be
\ba
F_{\hat S(\xi_m)\ket C}^\star(\alpha)&=(1-|t_{r_m}|^2)^{1/4}P_C(c_{r_m}\alpha+s_{r_m}e^{i\theta_m}\partial_\alpha)\cdot e^{-\frac12t_{r_m}e^{-i\theta_m}\alpha^2}\\
&=\frac1{\sqrt{c_{r_m}}}\sum_{n=0}^K{\frac{p_n}{\sqrt{n!}}(c_{r_m}\alpha+s_{r_m}e^{i\theta_m}\partial_\alpha)^n}\cdot e^{-\frac12t_{r_m}e^{-i\theta_m}\alpha^2}.
\ea
\label{FSC0}
\ee
where $t_{r_m}=\tanh(r_m)$.

The Hermite polynomials~\cite{abramowitz1965handbook} satisfy the following recurrence relation
\be
\mathit{He}_{n+1}(\alpha)=\alpha\mathit{He}_n(\alpha)-\partial_\alpha\mathit{He}_n(\alpha),
\label{recH}
\ee
for all $n\ge0$ and all $\alpha\in\mathbb C$, and $\mathit{He}_0=1$. Setting
\be
f_n(\alpha)=e^{\frac12t_{r_m}e^{-i\theta_m}\alpha^2}(c_{r_m}\alpha+s_{r_m}e^{i\theta_m}\partial_\alpha)^n\cdot e^{-\frac12t_{r_m}e^{-i\theta_m}\alpha^2},
\ee
we obtain $f_0(\alpha)=1$, and 
\be
\ba
f_{n+1}(\alpha)&=e^{\frac12t_{r_m}e^{-i\theta_m}\alpha^2}(c_{r_m}\alpha+s_{r_m}e^{i\theta_m}\partial_\alpha)\left[e^{-\frac12t_{r_m}e^{-i\theta_m}\alpha^2}f_n(\alpha)\right]\\
&=\frac\alpha{c_{r_m}}f_n(\alpha)+s_{r_m}e^{i\theta_m}\partial_\alpha f_n(\alpha).
\ea
\ee
Hence, with Eq.~(\ref{recH}), for all $n\ge0$ and all $\alpha\in\mathbb C$,
\be
f_n(\alpha)=\lambda_m^{n/2}\mathit{He}_n\left(\frac\alpha{c_{r_m}\sqrt{\lambda_m}}\right),
\ee
where we have set $\lambda_m=-e^{i\theta_m}t_{r_m}$. With Eq.~(\ref{FSC0}) we thus obtain
\be
\ba
F_{\hat S(\xi_m)\ket C}^\star(\alpha)&=\frac1{\sqrt{c_{r_m}}}\sum_{n=0}^K{\frac{p_n}{\sqrt{n!}}f_n(\alpha)}\cdot e^{-\frac12t_{r_m}e^{-i\theta_m}\alpha^2}\\
&=\frac1{\sqrt{c_{r_m}}}\sum_{n=0}^K{\frac{p_n\lambda_m^{n/2}}{\sqrt{n!}}\mathit{He}_n\left(\frac\alpha{c_{r_m}\sqrt{\lambda_m}}\right)}e^{-\frac12t_{r_m}e^{-i\theta_m}\alpha^2}.
\label{FSC}
\ea
\ee
From this calculation and Eq.~(\ref{QGmC1}) we deduce
\be
\ba
Q_{\hat G_m\ket C}(\alpha)&=\frac{e^{-|\alpha-\gamma_m|^2}}{\pi c_{r_m}}\left|\sum_{n=0}^K{\frac{p_n\lambda_m^{n/2}}{\sqrt{n!}}\mathit{He}_n\left(\frac{\alpha^*-\gamma_m^*}{c_{r_m}\sqrt{\lambda_m}}\right)}e^{-\frac12t_{r_m}e^{-i\theta_m}(\alpha^*-\gamma_m^*)^2}\right|^2\\
&\le\frac{e^{-|\alpha-\gamma_m|^2}}{\pi c_{r_m}}\left|e^{-\frac12t_{r_m}e^{-i\theta_m}(\alpha^*-\gamma_m^*)^2}\right|^2\sum_{n=0}^K{\frac{|p_n|^2}{n!}}\cdot\sum_{n=0}^K{\left|\lambda_m^{n/2}\mathit{He}_n\left(\frac{\alpha^*-\gamma_m^*}{c_{r_m}\sqrt{\lambda_m}}\right)\right|^2}\\
&=\frac1{\pi c_{r_m}}e^{-|\alpha-\gamma_m|^2-\frac12t_{r_m}[e^{i\theta_m}(\alpha-\gamma_m)^2+e^{-i\theta_m}(\alpha^*-\gamma_m^*)^2]}\sum_{n=0}^K{\left|t_{r_m}^{n/2}\mathit{He}_n\left(\frac{\alpha-\gamma_m}{c_{r_m}\sqrt{\lambda_m^*}}\right)\right|^2},
\ea
\ee
where we used Cauchy-Schwarz inequality in the second line, $|\lambda_m|=t_{r_m}$ and the fact that the coefficients of $\mathit{He}_n$ are real in the third line. Setting
\be
z_m(\alpha)=-\frac{ie^{\frac12i\theta_m}}{c_{r_m}}(\alpha-\gamma_m),
\label{zm}
\ee
for all $m\in\mathbb N$ and for all $\alpha\in\mathbb C$, we obtain
\be
\ba
Q_{\hat G_m\ket C}(\alpha)&\le\frac1{\pi c_{r_m}}e^{-|\alpha-\gamma_m|^2-\frac12t_{r_m}[e^{i\theta_m}(\alpha-\gamma_m)^2+e^{-i\theta_m}(\alpha^*-\gamma_m^*)^2]}\sum_{n=0}^K{\left|t_{r_m}^{n/2}\mathit{He}_n\left(\frac{e^{-i\theta_m}z_m(\alpha)}{\sqrt{t_{r_m}}}\right)\right|^2}\\
&=\frac1{\pi c_{r_m}}e^{-c_{r_m}^2|z_m(\alpha)|^2+\frac12c_{r_m}s_{r_m}[z_m^2(\alpha)+z_m^{*2}(\alpha)]}\sum_{n=0}^K{\left|t_{r_m}^{n/2}\mathit{He}_n\left(\frac{e^{-i\theta_m}z_m(\alpha)}{\sqrt{t_{r_m}}}\right)\right|^2}\\
&=\frac1{\pi c_{r_m}}e^{-c_{r_m}(c_{r_m}-s_{r_m})x_m^2(\alpha)}e^{-c_{r_m}(c_{r_m}+s_{r_m})y_m^2(\alpha)}\sum_{n=0}^K{\left|t_{r_m}^{n/2}\mathit{He}_n\left(\frac{e^{-i\theta_m}z_m(\alpha)}{\sqrt{t_{r_m}}}\right)\right|^2},
\ea
\label{boundQ1}
\ee
where $z_m(\alpha)=x_m(\alpha)+iy_m(\alpha)$. For all $r\in\mathbb R$,
\be
c_r(c_r-s_r)=\frac12(1+e^{-2r})>\frac12,
\ee
and
\be
c_r(c_r+s_r)=\frac12(1+e^{2r})>\frac12,
\ee
so with Eq.~(\ref{boundQ1}) we obtain
\be
Q_{\hat G_m\ket C}(\alpha)\le\frac1{\pi c_{r_m}}e^{-\frac12|z_m(\alpha)|^2}\sum_{n=0}^K{\left|t_{r_m}^{n/2}\mathit{He}_n\left(\frac{e^{-i\theta_m}z_m(\alpha)}{\sqrt{t_{r_m}}}\right)\right|^2}.
\label{boundQ2}
\ee
Finally, we obtain the following bound for all $n\in\{0,\dots,K\}$:
\be
\ba
\left|t_{r_m}^{n/2}\mathit{He}_n\left(\frac{e^{-i\theta_m}z_m(\alpha)}{\sqrt{t_{r_m}}}\right)\right|&=\left|t_{r_m}^{n/2}\sum_{k=0}^{\lfloor\frac n2\rfloor}{\frac{(-1)^kn!}{2^kk!(n-2k)!}\left(\frac{e^{-i\theta_m}z_m(\alpha)}{\sqrt{t_{r_m}}}\right)^{n-2k}}\right|\\
&\le\sum_{k=0}^{\lfloor\frac n2\rfloor}{\frac{n!}{2^kk!(n-2k)!}t_{r_m}^k|z_m(\alpha)|^{n-2k}}\\
&\le\sum_{k=0}^{\lfloor\frac n2\rfloor}{\frac{n!}{2^kk!(n-2k)!}|z_m(\alpha)|^{n-2k}},
\ea
\label{boundHe}
\ee
for all $m\in\mathbb N$ and all $\alpha\in\mathbb C$. Let us define for brevity the polynomial
\be
T(X):=\sum_{n=0}^K{\left(\sum_{k=0}^{\lfloor\frac n2\rfloor}{\frac{n!}{2^kk!(n-2k)!}X^{n-2k}}\right)^2}.
\ee
Plugging Eq.~(\ref{boundHe}) in Eq.~(\ref{boundQ2}) yields
\be
Q_{\hat G_m\ket C}(\alpha)\le\frac1{\pi c_{r_m}}e^{-\frac12|z_m(\alpha)|^2}T(|z_m(\alpha)|),
\label{boundQ3}
\ee
for all $m\in\mathbb N$ and all $\alpha\in\mathbb C$.

With this bound on the $Q$-function obtained, we can now prove that the sequences $\{\xi_m\}_{m\in\mathbb N}=\{r_me^{i\theta_m}\}_{m\in\mathbb N}$ and $\{\gamma_m\}_{m\in\mathbb N}$ are bounded.

Assuming that $\{r_m\}_{m\in\mathbb N}$ is unbounded implies that it has a subsequence $\{r_{m_k}\}_{k\in\mathbb N}$ going to infinity. Since the function $x\mapsto e^{-\frac12x^2}T(x)$ is bounded, $Q_{\hat G_{m_k}\ket C}(\alpha)\rightarrow0$ for all $\alpha\in\mathbb C$ when $k\rightarrow+\infty$ by Eq.~(\ref{boundQ3}). But $Q_{\hat G_{m_k}\ket C}(\alpha)\rightarrow Q_\psi(\alpha)$ for all $\alpha\in\mathbb C$ when $k\rightarrow+\infty$, by property of the convergence in trace norm. This would imply $Q_\psi(\alpha)=0$ for all $\alpha\in\mathbb C$, which is impossible since $\ket\psi$ is normalised. Hence $\{r_m\}_{m\in\mathbb N}$ is a bounded sequence, and so is $\{\xi_m\}_{m\in\mathbb N}$.

With the same reasoning, if $\{|z_m(\alpha)|\}_{m\in\mathbb N}$ was unbounded for all $\alpha\in\mathbb C$, this would imply by Eq.~(\ref{boundQ3}) that $Q_\psi(\alpha)=0$ for all $\alpha\in\mathbb C$, giving the same contradiction. Hence, there exists $\alpha_0\in\mathbb C$ such that the sequence $\{|z_m(\alpha_0)|\}_{m\in\mathbb N}$ is bounded. By Eq.~(\ref{zm}), this implies that the sequence $\{\gamma_m\}_{m\in\mathbb N}$ is also bounded, since the sequence $\{r_m\}_{m\in\mathbb N}$ is bounded.

\qed\\

The sequences $\{\xi_m\}_{m\in\mathbb N}$ and $\{\gamma_m\}_{m\in\mathbb N}$ being bounded, one can consider simultaneously converging subsequences $\{\xi_{m_k}\}_{k\in\mathbb N}$ and $\{\gamma_{m_k}\}_{k\in\mathbb N}$. We write $\xi=re^{i\theta}=\lim_{k\to\infty}{\xi_{m_k}}$ and $\gamma=\lim_{k\to\infty}{\gamma_{m_k}}$. On one hand, we have
\be
\ba
F_{\hat G_{m_k}\ket C}^\star(\alpha)&=F_{\hat D(\gamma_{m_k})\hat S(\xi_{m_k})\ket C}^\star(\alpha)\\
&=e^{\gamma_{m_k}\alpha-\frac12|\gamma_{m_k}|^2}F_{\hat S(\xi_{m_k})\ket C}^\star(\alpha-\gamma_{m_k}^*)\\
&=\frac1{\sqrt{c_{r_{m_k}}}}\sum_{n=0}^K{\frac{p_n\lambda_{m_k}^{n/2}}{\sqrt{n!}}\mathit{He}_n\left(\frac{\alpha-\gamma_{m_k}^*}{c_{r_{m_k}}\sqrt{\lambda_{m_k}}}\right)}e^{-\frac12t_{r_{m_k}}e^{-i\theta_{m_k}}(\alpha-\gamma_{m_k}^*)^2+\gamma_{m_k}\alpha-\frac12|\gamma_{m_k}|^2},
\ea
\label{compF}
\ee
for all $k\in\mathbb N$ and all $\alpha\in\mathbb C$, where we have used Eq.~(\ref{Fdisplaced}) in the second line, where $\lambda_{m_k}=-e^{i\theta_{m_k}}t_{r_{m_k}}$, and where we have used Eq.~(\ref{FSC}) in the last line. Setting $\lambda=-e^{i\theta}t_r$, we obtain
\be
\ba
\lim_{k\to\infty}{F_{\hat G_{m_k}\ket C}^\star(\alpha)}&=\frac1{\sqrt{c_r}}\sum_{n=0}^K{\frac{p_n\lambda^{n/2}}{\sqrt{n!}}\mathit{He}_n\left(\frac{\alpha-\gamma^*}{c_r\sqrt{\lambda}}\right)}e^{-\frac12t_re^{-i\theta}(\alpha-\gamma^*)^2+\gamma\alpha-\frac12|\gamma|^2}\\
&=F^\star_{\hat G\ket C}(\alpha),
\ea
\label{limF1}
\ee
for all $\alpha\in\mathbb C$, where $\hat G=\hat D(\gamma)\hat S(\xi)$, and where the second line comes from reversing the calculations of Eq.~(\ref{compF}). On the other hand, for all $\alpha\in\mathbb C$,
\be
\ba
\lim_{k\to\infty}{F_{\hat G_{m_k}\ket C}^\star(\alpha)}&=e^{\frac12|\alpha|^2}\lim_{k\to\infty}{\braket{\alpha^*|\hat G_{m_k}|C}}\\
&=e^{\frac12|\alpha|^2}\braket{\alpha^*|\psi}\\
&=F_\psi^\star(\alpha),
\ea
\label{limF2}
\ee
by property of the convergence in trace norm. 
Combining Eq.~(\ref{limF1}) and Eq.~(\ref{limF2}) yields
\be
F_\psi^\star(\alpha)=F_{\hat G\ket C}^\star(\alpha),
\ee
for all $\alpha\in\mathbb C$. By Lemma~\ref{unique}, this implies that $\ket\psi=\hat G\ket C\in R_K$. This shows that $\overline{\bigcup_{K=0}^N{R_K}}=\bigcup_{K=0}^N{R_K}$, so $\overline{R_N}\subset\bigcup_{K=0}^N{R_K}$, which concludes the proof.

\end{proof}


\section{Density of the set of states of finite stellar rank.}
\label{app:dense}

\noindent We prove in this section Lemma~\ref{dense} from the main text:

\medskip

\textit{The set of states of finite stellar rank is dense for the trace norm in the set of normalised pure single-mode states:}
\be
\overline{\underset{N\in\mathbb N}{\bigcup}{R_N}}=\mathcal H_\infty,
\ee
\textit{where $\overline X$ denotes the closure of $X$ for the trace norm in the set of normalised states $\mathcal H_\infty$.}

\begin{proof}

Recall that the set of normalised pure single-mode states is closed for the trace norm in the whole Hilbert space, since it is the reciprocal image of $\{1\}$ by the trace norm, which is Lipschitz continuous with Lipschitz constant $1$, hence continuous.

Let $\ket\psi\in\mathcal H_\infty$ be a normalised state. We consider the sequence of normalised cut-off states
\be
\ket{\psi_m}=\frac1{\mathcal{N}_m^{1/2}}\sum_{n=0}^m{\psi_n\ket n},
\ee
where $\mathcal{N}_m=\sum_{n=0}^m{|\psi_n|^2}$ is a normalising factor (non-zero for $m$ large enough). All the states $\ket{\psi_m}$ have a finite support over the Fock basis, so their stellar function is a polynomial. Hence $\{\ket{\psi_m}\}_{m\in\mathbb N}\in\bigcup_{N\in\mathbb N}{R_N}$.

Moreover, for all $m\in\mathbb N$,
\be
\ba
D_1(\psi_m,\psi)&=\sqrt{1-|\braket{\psi_m|\psi}|^2}\\
&=\sqrt{1-\sum_{n=0}^m{|\psi_n|^2}}\\
&=\sqrt{\sum_{n\ge m+1}{|\psi_n|^2}},
\ea
\ee
where we used that $\ket\psi$ and $\ket{\psi_m}$ are pure states in the first line, and the fact that $\ket\psi$ is normalised in the third line. Moreover, $\sum_{n\ge m+1}{|\psi_n|^2}\rightarrow0$ when $m\rightarrow+\infty$, because $\ket\psi$ is normalised. Hence, $\{\ket{\psi_m}\}_{m\in\mathbb N}$ converges in trace norm to $\ket\psi$, which concludes the proof.

\end{proof}


\bibliographystyle{apsrev}
\bibliography{bibliography}

\begin{thebibliography}{63}
\expandafter\ifx\csname natexlab\endcsname\relax\def\natexlab#1{#1}\fi
\expandafter\ifx\csname bibnamefont\endcsname\relax
  \def\bibnamefont#1{#1}\fi
\expandafter\ifx\csname bibfnamefont\endcsname\relax
  \def\bibfnamefont#1{#1}\fi
\expandafter\ifx\csname citenamefont\endcsname\relax
  \def\citenamefont#1{#1}\fi
\expandafter\ifx\csname url\endcsname\relax
  \def\url#1{\texttt{#1}}\fi
\expandafter\ifx\csname urlprefix\endcsname\relax\def\urlprefix{URL }\fi
\providecommand{\bibinfo}[2]{#2}
\providecommand{\eprint}[2][]{\url{#2}}

\bibitem[{\citenamefont{Bennett et~al.}(1993)\citenamefont{Bennett, Brassard,
  Cr{\'e}peau, Jozsa, Peres, and Wootters}}]{bennett1993teleporting}
\bibinfo{author}{\bibfnamefont{C.~H.} \bibnamefont{Bennett}},
  \bibinfo{author}{\bibfnamefont{G.}~\bibnamefont{Brassard}},
  \bibinfo{author}{\bibfnamefont{C.}~\bibnamefont{Cr{\'e}peau}},
  \bibinfo{author}{\bibfnamefont{R.}~\bibnamefont{Jozsa}},
  \bibinfo{author}{\bibfnamefont{A.}~\bibnamefont{Peres}}, \bibnamefont{and}
  \bibinfo{author}{\bibfnamefont{W.~K.} \bibnamefont{Wootters}},
  \bibinfo{journal}{Physical review letters} \textbf{\bibinfo{volume}{70}},
  \bibinfo{pages}{1895} (\bibinfo{year}{1993}).

\bibitem[{\citenamefont{Shor}(1994)}]{shor1994algorithms}
\bibinfo{author}{\bibfnamefont{P.~W.} \bibnamefont{Shor}}, in
  \emph{\bibinfo{booktitle}{Proceedings 35th annual symposium on foundations of
  computer science}} (\bibinfo{organization}{Ieee}, \bibinfo{year}{1994}), pp.
  \bibinfo{pages}{124--134}.

\bibitem[{\citenamefont{Braunstein and van Loock}(2005)}]{Braunstein2005}
\bibinfo{author}{\bibfnamefont{S.~L.} \bibnamefont{Braunstein}}
  \bibnamefont{and} \bibinfo{author}{\bibfnamefont{P.}~\bibnamefont{van
  Loock}}, \bibinfo{journal}{Rev. Mod. Phys.} \textbf{\bibinfo{volume}{77}},
  \bibinfo{pages}{513} (\bibinfo{year}{2005}).

\bibitem[{\citenamefont{Yokoyama et~al.}(2013)\citenamefont{Yokoyama, Ukai,
  Armstrong, Sornphiphatphong, Kaji, Suzuki, Yoshikawa, Yonezawa, Menicucci,
  and Furusawa}}]{yokoyama2013ultra}
\bibinfo{author}{\bibfnamefont{S.}~\bibnamefont{Yokoyama}},
  \bibinfo{author}{\bibfnamefont{R.}~\bibnamefont{Ukai}},
  \bibinfo{author}{\bibfnamefont{S.~C.} \bibnamefont{Armstrong}},
  \bibinfo{author}{\bibfnamefont{C.}~\bibnamefont{Sornphiphatphong}},
  \bibinfo{author}{\bibfnamefont{T.}~\bibnamefont{Kaji}},
  \bibinfo{author}{\bibfnamefont{S.}~\bibnamefont{Suzuki}},
  \bibinfo{author}{\bibfnamefont{J.-i.} \bibnamefont{Yoshikawa}},
  \bibinfo{author}{\bibfnamefont{H.}~\bibnamefont{Yonezawa}},
  \bibinfo{author}{\bibfnamefont{N.~C.} \bibnamefont{Menicucci}},
  \bibnamefont{and} \bibinfo{author}{\bibfnamefont{A.}~\bibnamefont{Furusawa}},
  \bibinfo{journal}{Nature Photonics} \textbf{\bibinfo{volume}{7}},
  \bibinfo{pages}{982} (\bibinfo{year}{2013}).

\bibitem[{\citenamefont{Yoshikawa et~al.}(2016)\citenamefont{Yoshikawa,
  Yokoyama, Kaji, Sornphiphatphong, Shiozawa, Makino, and
  Furusawa}}]{yoshikawa2016invited}
\bibinfo{author}{\bibfnamefont{J.-i.} \bibnamefont{Yoshikawa}},
  \bibinfo{author}{\bibfnamefont{S.}~\bibnamefont{Yokoyama}},
  \bibinfo{author}{\bibfnamefont{T.}~\bibnamefont{Kaji}},
  \bibinfo{author}{\bibfnamefont{C.}~\bibnamefont{Sornphiphatphong}},
  \bibinfo{author}{\bibfnamefont{Y.}~\bibnamefont{Shiozawa}},
  \bibinfo{author}{\bibfnamefont{K.}~\bibnamefont{Makino}}, \bibnamefont{and}
  \bibinfo{author}{\bibfnamefont{A.}~\bibnamefont{Furusawa}},
  \bibinfo{journal}{APL Photonics} \textbf{\bibinfo{volume}{1}},
  \bibinfo{pages}{060801} (\bibinfo{year}{2016}).

\bibitem[{\citenamefont{Cahill and Glauber}(1969)}]{cahill1969density}
\bibinfo{author}{\bibfnamefont{K.~E.} \bibnamefont{Cahill}} \bibnamefont{and}
  \bibinfo{author}{\bibfnamefont{R.~J.} \bibnamefont{Glauber}},
  \bibinfo{journal}{Physical Review} \textbf{\bibinfo{volume}{177}},
  \bibinfo{pages}{1882} (\bibinfo{year}{1969}).

\bibitem[{\citenamefont{Marian and Marian}(1993)}]{marian1993squeezed}
\bibinfo{author}{\bibfnamefont{P.}~\bibnamefont{Marian}} \bibnamefont{and}
  \bibinfo{author}{\bibfnamefont{T.~A.} \bibnamefont{Marian}},
  \bibinfo{journal}{Physical Review A} \textbf{\bibinfo{volume}{47}},
  \bibinfo{pages}{4474} (\bibinfo{year}{1993}).

\bibitem[{\citenamefont{Ferraro et~al.}(2005)\citenamefont{Ferraro, Olivares,
  and Paris}}]{ferraro2005gaussian}
\bibinfo{author}{\bibfnamefont{A.}~\bibnamefont{Ferraro}},
  \bibinfo{author}{\bibfnamefont{S.}~\bibnamefont{Olivares}}, \bibnamefont{and}
  \bibinfo{author}{\bibfnamefont{M.~G.~A.} \bibnamefont{Paris}},
  \bibinfo{journal}{arXiv preprint quant-ph/0503237}  (\bibinfo{year}{2005}).

\bibitem[{\citenamefont{Weedbrook et~al.}(2012)\citenamefont{Weedbrook,
  Pirandola, Garc\'{i}a-Patr\'{o}n, Cerf, Ralph, Shapiro, and
  Lloyd}}]{Lloyd2012}
\bibinfo{author}{\bibfnamefont{C.}~\bibnamefont{Weedbrook}},
  \bibinfo{author}{\bibfnamefont{S.}~\bibnamefont{Pirandola}},
  \bibinfo{author}{\bibfnamefont{R.}~\bibnamefont{Garc\'{i}a-Patr\'{o}n}},
  \bibinfo{author}{\bibfnamefont{N.~J.} \bibnamefont{Cerf}},
  \bibinfo{author}{\bibfnamefont{T.~C.} \bibnamefont{Ralph}},
  \bibinfo{author}{\bibfnamefont{J.~H.} \bibnamefont{Shapiro}},
  \bibnamefont{and} \bibinfo{author}{\bibfnamefont{S.}~\bibnamefont{Lloyd}},
  \bibinfo{journal}{Rev. Mod. Phys.} \textbf{\bibinfo{volume}{84}},
  \bibinfo{pages}{621} (\bibinfo{year}{2012}).

\bibitem[{\citenamefont{Adesso et~al.}(2014)\citenamefont{Adesso, Ragy, and
  Lee}}]{adesso2014continuous}
\bibinfo{author}{\bibfnamefont{G.}~\bibnamefont{Adesso}},
  \bibinfo{author}{\bibfnamefont{S.}~\bibnamefont{Ragy}}, \bibnamefont{and}
  \bibinfo{author}{\bibfnamefont{A.~R.} \bibnamefont{Lee}},
  \bibinfo{journal}{Open Systems \& Information Dynamics}
  \textbf{\bibinfo{volume}{21}}, \bibinfo{pages}{1440001}
  (\bibinfo{year}{2014}).

\bibitem[{\citenamefont{Mari and Eisert}(2012)}]{mari2012positive}
\bibinfo{author}{\bibfnamefont{A.}~\bibnamefont{Mari}} \bibnamefont{and}
  \bibinfo{author}{\bibfnamefont{J.}~\bibnamefont{Eisert}},
  \bibinfo{journal}{Physical review letters} \textbf{\bibinfo{volume}{109}},
  \bibinfo{pages}{230503} (\bibinfo{year}{2012}).

\bibitem[{\citenamefont{Gottesman et~al.}(2001)\citenamefont{Gottesman, Kitaev,
  and Preskill}}]{Gottesman2001}
\bibinfo{author}{\bibfnamefont{D.}~\bibnamefont{Gottesman}},
  \bibinfo{author}{\bibfnamefont{A.}~\bibnamefont{Kitaev}}, \bibnamefont{and}
  \bibinfo{author}{\bibfnamefont{J.}~\bibnamefont{Preskill}},
  \bibinfo{journal}{Phys. Rev. A} \textbf{\bibinfo{volume}{64}},
  \bibinfo{pages}{012310} (\bibinfo{year}{2001}).

\bibitem[{\citenamefont{Baragiola et~al.}(2019)\citenamefont{Baragiola,
  Pantaleoni, Alexander, Karanjai, and Menicucci}}]{baragiola2019all}
\bibinfo{author}{\bibfnamefont{B.~Q.} \bibnamefont{Baragiola}},
  \bibinfo{author}{\bibfnamefont{G.}~\bibnamefont{Pantaleoni}},
  \bibinfo{author}{\bibfnamefont{R.~N.} \bibnamefont{Alexander}},
  \bibinfo{author}{\bibfnamefont{A.}~\bibnamefont{Karanjai}}, \bibnamefont{and}
  \bibinfo{author}{\bibfnamefont{N.~C.} \bibnamefont{Menicucci}},
  \bibinfo{journal}{arXiv preprint arXiv:1903.00012}  (\bibinfo{year}{2019}).

\bibitem[{\citenamefont{Ghose and Sanders}(2007)}]{ghose2007non}
\bibinfo{author}{\bibfnamefont{S.}~\bibnamefont{Ghose}} \bibnamefont{and}
  \bibinfo{author}{\bibfnamefont{B.~C.} \bibnamefont{Sanders}},
  \bibinfo{journal}{Journal of Modern Optics} \textbf{\bibinfo{volume}{54}},
  \bibinfo{pages}{855} (\bibinfo{year}{2007}).

\bibitem[{\citenamefont{Eisert et~al.}(2002)\citenamefont{Eisert, Scheel, and
  Plenio}}]{eisert2002distilling}
\bibinfo{author}{\bibfnamefont{J.}~\bibnamefont{Eisert}},
  \bibinfo{author}{\bibfnamefont{S.}~\bibnamefont{Scheel}}, \bibnamefont{and}
  \bibinfo{author}{\bibfnamefont{M.~B.} \bibnamefont{Plenio}},
  \bibinfo{journal}{Physical review letters} \textbf{\bibinfo{volume}{89}},
  \bibinfo{pages}{137903} (\bibinfo{year}{2002}).

\bibitem[{\citenamefont{Fiur{\'a}{\v{s}}ek}(2002)}]{fiuravsek2002gaussian}
\bibinfo{author}{\bibfnamefont{J.}~\bibnamefont{Fiur{\'a}{\v{s}}ek}},
  \bibinfo{journal}{Physical review letters} \textbf{\bibinfo{volume}{89}},
  \bibinfo{pages}{137904} (\bibinfo{year}{2002}).

\bibitem[{\citenamefont{Giedke and Cirac}(2002)}]{giedke2002characterization}
\bibinfo{author}{\bibfnamefont{G.}~\bibnamefont{Giedke}} \bibnamefont{and}
  \bibinfo{author}{\bibfnamefont{J.~I.} \bibnamefont{Cirac}},
  \bibinfo{journal}{Physical Review A} \textbf{\bibinfo{volume}{66}},
  \bibinfo{pages}{032316} (\bibinfo{year}{2002}).

\bibitem[{\citenamefont{Wenger et~al.}(2003)\citenamefont{Wenger, Hafezi,
  Grosshans, Tualle-Brouri, and Grangier}}]{wenger2003maximal}
\bibinfo{author}{\bibfnamefont{J.}~\bibnamefont{Wenger}},
  \bibinfo{author}{\bibfnamefont{M.}~\bibnamefont{Hafezi}},
  \bibinfo{author}{\bibfnamefont{F.}~\bibnamefont{Grosshans}},
  \bibinfo{author}{\bibfnamefont{R.}~\bibnamefont{Tualle-Brouri}},
  \bibnamefont{and} \bibinfo{author}{\bibfnamefont{P.}~\bibnamefont{Grangier}},
  \bibinfo{journal}{Physical Review A} \textbf{\bibinfo{volume}{67}},
  \bibinfo{pages}{012105} (\bibinfo{year}{2003}).

\bibitem[{\citenamefont{Garc{\'\i}a-Patr{\'o}n
  et~al.}(2004)\citenamefont{Garc{\'\i}a-Patr{\'o}n, Fiur{\'a}{\v{s}}ek, Cerf,
  Wenger, Tualle-Brouri, and Grangier}}]{garcia2004proposal}
\bibinfo{author}{\bibfnamefont{R.}~\bibnamefont{Garc{\'\i}a-Patr{\'o}n}},
  \bibinfo{author}{\bibfnamefont{J.}~\bibnamefont{Fiur{\'a}{\v{s}}ek}},
  \bibinfo{author}{\bibfnamefont{N.~J.} \bibnamefont{Cerf}},
  \bibinfo{author}{\bibfnamefont{J.}~\bibnamefont{Wenger}},
  \bibinfo{author}{\bibfnamefont{R.}~\bibnamefont{Tualle-Brouri}},
  \bibnamefont{and} \bibinfo{author}{\bibfnamefont{P.}~\bibnamefont{Grangier}},
  \bibinfo{journal}{Physical review letters} \textbf{\bibinfo{volume}{93}},
  \bibinfo{pages}{130409} (\bibinfo{year}{2004}).

\bibitem[{\citenamefont{Niset et~al.}(2009)\citenamefont{Niset,
  Fiur{\'a}{\v{s}}ek, and Cerf}}]{niset2009no}
\bibinfo{author}{\bibfnamefont{J.}~\bibnamefont{Niset}},
  \bibinfo{author}{\bibfnamefont{J.}~\bibnamefont{Fiur{\'a}{\v{s}}ek}},
  \bibnamefont{and} \bibinfo{author}{\bibfnamefont{N.~J.} \bibnamefont{Cerf}},
  \bibinfo{journal}{Physical review letters} \textbf{\bibinfo{volume}{102}},
  \bibinfo{pages}{120501} (\bibinfo{year}{2009}).

\bibitem[{\citenamefont{Adesso et~al.}(2009)\citenamefont{Adesso, Dell'Anno,
  De~Siena, Illuminati, and Souza}}]{adesso2009optimal}
\bibinfo{author}{\bibfnamefont{G.}~\bibnamefont{Adesso}},
  \bibinfo{author}{\bibfnamefont{F.}~\bibnamefont{Dell'Anno}},
  \bibinfo{author}{\bibfnamefont{S.}~\bibnamefont{De~Siena}},
  \bibinfo{author}{\bibfnamefont{F.}~\bibnamefont{Illuminati}},
  \bibnamefont{and} \bibinfo{author}{\bibfnamefont{L.~A.~M.}
  \bibnamefont{Souza}}, \bibinfo{journal}{Physical Review A}
  \textbf{\bibinfo{volume}{79}}, \bibinfo{pages}{040305}
  (\bibinfo{year}{2009}).

\bibitem[{\citenamefont{Barbosa et~al.}(2019)\citenamefont{Barbosa, Douce,
  Emeriau, Kashefi, and Mansfield}}]{barbosa2019continuous}
\bibinfo{author}{\bibfnamefont{R.~S.} \bibnamefont{Barbosa}},
  \bibinfo{author}{\bibfnamefont{T.}~\bibnamefont{Douce}},
  \bibinfo{author}{\bibfnamefont{P.-E.} \bibnamefont{Emeriau}},
  \bibinfo{author}{\bibfnamefont{E.}~\bibnamefont{Kashefi}}, \bibnamefont{and}
  \bibinfo{author}{\bibfnamefont{S.}~\bibnamefont{Mansfield}},
  \bibinfo{journal}{arXiv preprint arXiv:1905.08267}  (\bibinfo{year}{2019}).

\bibitem[{\citenamefont{Takagi and Zhuang}(2018)}]{takagi2018convex}
\bibinfo{author}{\bibfnamefont{R.}~\bibnamefont{Takagi}} \bibnamefont{and}
  \bibinfo{author}{\bibfnamefont{Q.}~\bibnamefont{Zhuang}},
  \bibinfo{journal}{Physical Review A} \textbf{\bibinfo{volume}{97}},
  \bibinfo{pages}{062337} (\bibinfo{year}{2018}).

\bibitem[{\citenamefont{Zhuang et~al.}(2018)\citenamefont{Zhuang, Shor, and
  Shapiro}}]{zhuang2018resource}
\bibinfo{author}{\bibfnamefont{Q.}~\bibnamefont{Zhuang}},
  \bibinfo{author}{\bibfnamefont{P.~W.} \bibnamefont{Shor}}, \bibnamefont{and}
  \bibinfo{author}{\bibfnamefont{J.~H.} \bibnamefont{Shapiro}},
  \bibinfo{journal}{Physical Review A} \textbf{\bibinfo{volume}{97}},
  \bibinfo{pages}{052317} (\bibinfo{year}{2018}).

\bibitem[{\citenamefont{Albarelli et~al.}(2018)\citenamefont{Albarelli, Genoni,
  Paris, and Ferraro}}]{albarelli2018resource}
\bibinfo{author}{\bibfnamefont{F.}~\bibnamefont{Albarelli}},
  \bibinfo{author}{\bibfnamefont{M.~G.} \bibnamefont{Genoni}},
  \bibinfo{author}{\bibfnamefont{M.~G.~A.} \bibnamefont{Paris}},
  \bibnamefont{and} \bibinfo{author}{\bibfnamefont{A.}~\bibnamefont{Ferraro}},
  \bibinfo{journal}{Physical Review A} \textbf{\bibinfo{volume}{98}},
  \bibinfo{pages}{052350} (\bibinfo{year}{2018}).

\bibitem[{\citenamefont{Lami et~al.}(2018)\citenamefont{Lami, Regula, Wang,
  Nichols, Winter, and Adesso}}]{lami2018gaussian}
\bibinfo{author}{\bibfnamefont{L.}~\bibnamefont{Lami}},
  \bibinfo{author}{\bibfnamefont{B.}~\bibnamefont{Regula}},
  \bibinfo{author}{\bibfnamefont{X.}~\bibnamefont{Wang}},
  \bibinfo{author}{\bibfnamefont{R.}~\bibnamefont{Nichols}},
  \bibinfo{author}{\bibfnamefont{A.}~\bibnamefont{Winter}}, \bibnamefont{and}
  \bibinfo{author}{\bibfnamefont{G.}~\bibnamefont{Adesso}},
  \bibinfo{journal}{Physical Review A} \textbf{\bibinfo{volume}{98}},
  \bibinfo{pages}{022335} (\bibinfo{year}{2018}).

\bibitem[{\citenamefont{Hudson}(1974)}]{hudson1974wigner}
\bibinfo{author}{\bibfnamefont{R.~L.} \bibnamefont{Hudson}},
  \bibinfo{journal}{Reports on Mathematical Physics}
  \textbf{\bibinfo{volume}{6}}, \bibinfo{pages}{249} (\bibinfo{year}{1974}).

\bibitem[{\citenamefont{Soto and Claverie}(1983)}]{soto1983wigner}
\bibinfo{author}{\bibfnamefont{F.}~\bibnamefont{Soto}} \bibnamefont{and}
  \bibinfo{author}{\bibfnamefont{P.}~\bibnamefont{Claverie}},
  \bibinfo{journal}{Journal of Mathematical Physics}
  \textbf{\bibinfo{volume}{24}}, \bibinfo{pages}{97} (\bibinfo{year}{1983}).

\bibitem[{\citenamefont{Kenfack and
  {\.Z}yczkowski}(2004)}]{kenfack2004negativity}
\bibinfo{author}{\bibfnamefont{A.}~\bibnamefont{Kenfack}} \bibnamefont{and}
  \bibinfo{author}{\bibfnamefont{K.}~\bibnamefont{{\.Z}yczkowski}},
  \bibinfo{journal}{Journal of Optics B: Quantum and Semiclassical Optics}
  \textbf{\bibinfo{volume}{6}}, \bibinfo{pages}{396} (\bibinfo{year}{2004}).

\bibitem[{\citenamefont{Genoni et~al.}(2007)\citenamefont{Genoni, Paris, and
  Banaszek}}]{genoni2007measure}
\bibinfo{author}{\bibfnamefont{M.~G.} \bibnamefont{Genoni}},
  \bibinfo{author}{\bibfnamefont{M.~G.~A.} \bibnamefont{Paris}},
  \bibnamefont{and} \bibinfo{author}{\bibfnamefont{K.}~\bibnamefont{Banaszek}},
  \bibinfo{journal}{Physical Review A} \textbf{\bibinfo{volume}{76}},
  \bibinfo{pages}{042327} (\bibinfo{year}{2007}).

\bibitem[{\citenamefont{Filip and Mi{\v{s}}ta~Jr}(2011)}]{filip2011detecting}
\bibinfo{author}{\bibfnamefont{R.}~\bibnamefont{Filip}} \bibnamefont{and}
  \bibinfo{author}{\bibfnamefont{L.}~\bibnamefont{Mi{\v{s}}ta~Jr}},
  \bibinfo{journal}{Physical Review Letters} \textbf{\bibinfo{volume}{106}},
  \bibinfo{pages}{200401} (\bibinfo{year}{2011}).

\bibitem[{\citenamefont{Genoni et~al.}(2013)\citenamefont{Genoni, Palma,
  Tufarelli, Olivares, Kim, and Paris}}]{genoni2013detecting}
\bibinfo{author}{\bibfnamefont{M.~G.} \bibnamefont{Genoni}},
  \bibinfo{author}{\bibfnamefont{M.~L.} \bibnamefont{Palma}},
  \bibinfo{author}{\bibfnamefont{T.}~\bibnamefont{Tufarelli}},
  \bibinfo{author}{\bibfnamefont{S.}~\bibnamefont{Olivares}},
  \bibinfo{author}{\bibfnamefont{M.~S.} \bibnamefont{Kim}}, \bibnamefont{and}
  \bibinfo{author}{\bibfnamefont{M.~G.~A.} \bibnamefont{Paris}},
  \bibinfo{journal}{Physical Review A} \textbf{\bibinfo{volume}{87}},
  \bibinfo{pages}{062104} (\bibinfo{year}{2013}).

\bibitem[{\citenamefont{Hughes et~al.}(2014)\citenamefont{Hughes, Genoni,
  Tufarelli, Paris, and Kim}}]{hughes2014quantum}
\bibinfo{author}{\bibfnamefont{C.}~\bibnamefont{Hughes}},
  \bibinfo{author}{\bibfnamefont{M.~G.} \bibnamefont{Genoni}},
  \bibinfo{author}{\bibfnamefont{T.}~\bibnamefont{Tufarelli}},
  \bibinfo{author}{\bibfnamefont{M.~G.~A.} \bibnamefont{Paris}},
  \bibnamefont{and} \bibinfo{author}{\bibfnamefont{M.~S.} \bibnamefont{Kim}},
  \bibinfo{journal}{Physical Review A} \textbf{\bibinfo{volume}{90}},
  \bibinfo{pages}{013810} (\bibinfo{year}{2014}).

\bibitem[{\citenamefont{L{\"u}tkenhaus and
  Barnett}(1995)}]{lutkenhaus1995nonclassical}
\bibinfo{author}{\bibfnamefont{N.}~\bibnamefont{L{\"u}tkenhaus}}
  \bibnamefont{and} \bibinfo{author}{\bibfnamefont{S.~M.}
  \bibnamefont{Barnett}}, \bibinfo{journal}{Physical Review A}
  \textbf{\bibinfo{volume}{51}}, \bibinfo{pages}{3340} (\bibinfo{year}{1995}).

\bibitem[{\citenamefont{Bargmann}(1961)}]{bargmann1961hilbert}
\bibinfo{author}{\bibfnamefont{V.}~\bibnamefont{Bargmann}},
  \bibinfo{journal}{Communications on pure and applied mathematics}
  \textbf{\bibinfo{volume}{14}}, \bibinfo{pages}{187} (\bibinfo{year}{1961}).

\bibitem[{\citenamefont{Segal and Mackey}(1963)}]{segal1963mathematical}
\bibinfo{author}{\bibfnamefont{I.~E.} \bibnamefont{Segal}} \bibnamefont{and}
  \bibinfo{author}{\bibfnamefont{G.~W.} \bibnamefont{Mackey}},
  \emph{\bibinfo{title}{Mathematical problems of relativistic physics}},
  vol.~\bibinfo{volume}{2} (\bibinfo{publisher}{American Mathematical Soc.},
  \bibinfo{year}{1963}).

\bibitem[{\citenamefont{Lachman et~al.}(2018)\citenamefont{Lachman, Straka,
  Hlou{\v{s}}ek, Je{\v{z}}ek, and Filip}}]{lachman2018faithful}
\bibinfo{author}{\bibfnamefont{L.}~\bibnamefont{Lachman}},
  \bibinfo{author}{\bibfnamefont{I.}~\bibnamefont{Straka}},
  \bibinfo{author}{\bibfnamefont{J.}~\bibnamefont{Hlou{\v{s}}ek}},
  \bibinfo{author}{\bibfnamefont{M.}~\bibnamefont{Je{\v{z}}ek}},
  \bibnamefont{and} \bibinfo{author}{\bibfnamefont{R.}~\bibnamefont{Filip}},
  \bibinfo{journal}{arXiv preprint arXiv:1810.02546}  (\bibinfo{year}{2018}).

\bibitem[{\citenamefont{Leboeuf and Voros}(1990)}]{leboeuf1990chaos}
\bibinfo{author}{\bibfnamefont{P.}~\bibnamefont{Leboeuf}} \bibnamefont{and}
  \bibinfo{author}{\bibfnamefont{A.}~\bibnamefont{Voros}},
  \bibinfo{journal}{Journal of Physics A: Mathematical and General}
  \textbf{\bibinfo{volume}{23}}, \bibinfo{pages}{1765} (\bibinfo{year}{1990}).

\bibitem[{\citenamefont{Arranz et~al.}(1996)\citenamefont{Arranz, Borondo, and
  Benito}}]{arranz1996distribution}
\bibinfo{author}{\bibfnamefont{F.}~\bibnamefont{Arranz}},
  \bibinfo{author}{\bibfnamefont{F.}~\bibnamefont{Borondo}}, \bibnamefont{and}
  \bibinfo{author}{\bibfnamefont{R.~M.} \bibnamefont{Benito}},
  \bibinfo{journal}{Physical Review E} \textbf{\bibinfo{volume}{54}},
  \bibinfo{pages}{2458} (\bibinfo{year}{1996}).

\bibitem[{\citenamefont{Korsch et~al.}(1997)\citenamefont{Korsch, M{\"u}ller,
  and Wiescher}}]{korsch1997zeros}
\bibinfo{author}{\bibfnamefont{H.}~\bibnamefont{Korsch}},
  \bibinfo{author}{\bibfnamefont{C.}~\bibnamefont{M{\"u}ller}},
  \bibnamefont{and} \bibinfo{author}{\bibfnamefont{H.}~\bibnamefont{Wiescher}},
  \bibinfo{journal}{Journal of Physics A: Mathematical and General}
  \textbf{\bibinfo{volume}{30}}, \bibinfo{pages}{L677} (\bibinfo{year}{1997}).

\bibitem[{\citenamefont{Biswas and Sinha}(1999)}]{biswas1999distribution}
\bibinfo{author}{\bibfnamefont{D.}~\bibnamefont{Biswas}} \bibnamefont{and}
  \bibinfo{author}{\bibfnamefont{S.}~\bibnamefont{Sinha}},
  \bibinfo{journal}{Physical Review E} \textbf{\bibinfo{volume}{60}},
  \bibinfo{pages}{408} (\bibinfo{year}{1999}).

\bibitem[{\citenamefont{Perelomov}(1971)}]{perelomov1971completeness}
\bibinfo{author}{\bibfnamefont{A.~M.} \bibnamefont{Perelomov}},
  \bibinfo{journal}{Theoretical and Mathematical Physics}
  \textbf{\bibinfo{volume}{6}}, \bibinfo{pages}{156} (\bibinfo{year}{1971}).

\bibitem[{\citenamefont{Bacry et~al.}(1975)\citenamefont{Bacry, Grossmann, and
  Zak}}]{bacry1975proof}
\bibinfo{author}{\bibfnamefont{H.}~\bibnamefont{Bacry}},
  \bibinfo{author}{\bibfnamefont{A.}~\bibnamefont{Grossmann}},
  \bibnamefont{and} \bibinfo{author}{\bibfnamefont{J.}~\bibnamefont{Zak}},
  \bibinfo{journal}{Physical Review B} \textbf{\bibinfo{volume}{12}},
  \bibinfo{pages}{1118} (\bibinfo{year}{1975}).

\bibitem[{\citenamefont{Boon and Zak}(1978)}]{boon1978discrete}
\bibinfo{author}{\bibfnamefont{M.}~\bibnamefont{Boon}} \bibnamefont{and}
  \bibinfo{author}{\bibfnamefont{J.}~\bibnamefont{Zak}},
  \bibinfo{journal}{Physical Review B} \textbf{\bibinfo{volume}{18}},
  \bibinfo{pages}{6744} (\bibinfo{year}{1978}).

\bibitem[{\citenamefont{Vourdas}(2006)}]{vourdas2006analytic}
\bibinfo{author}{\bibfnamefont{A.}~\bibnamefont{Vourdas}},
  \bibinfo{journal}{Journal of Physics A: Mathematical and General}
  \textbf{\bibinfo{volume}{39}}, \bibinfo{pages}{R65} (\bibinfo{year}{2006}).

\bibitem[{\citenamefont{Gagatsos and Guha}(2019)}]{PhysRevA.99.053816}
\bibinfo{author}{\bibfnamefont{C.~N.} \bibnamefont{Gagatsos}} \bibnamefont{and}
  \bibinfo{author}{\bibfnamefont{S.}~\bibnamefont{Guha}},
  \bibinfo{journal}{Phys. Rev. A} \textbf{\bibinfo{volume}{99}},
  \bibinfo{pages}{053816} (\bibinfo{year}{2019}).

\bibitem[{\citenamefont{Sidoli and Berggren}(2007)}]{sidoli2007arabic}
\bibinfo{author}{\bibfnamefont{N.}~\bibnamefont{Sidoli}} \bibnamefont{and}
  \bibinfo{author}{\bibfnamefont{J.~L.} \bibnamefont{Berggren}},
  \bibinfo{journal}{Sciamvs} \textbf{\bibinfo{volume}{8}}, \bibinfo{pages}{37}
  (\bibinfo{year}{2007}).

\bibitem[{\citenamefont{Tualle and Voros}(1995)}]{tualle1995normal}
\bibinfo{author}{\bibfnamefont{J.-M.} \bibnamefont{Tualle}} \bibnamefont{and}
  \bibinfo{author}{\bibfnamefont{A.}~\bibnamefont{Voros}},
  \bibinfo{journal}{Chaos, Solitons \& Fractals} \textbf{\bibinfo{volume}{5}},
  \bibinfo{pages}{1085} (\bibinfo{year}{1995}).

\bibitem[{\citenamefont{Terhal and Horodecki}(2000)}]{terhal2000schmidt}
\bibinfo{author}{\bibfnamefont{B.~M.} \bibnamefont{Terhal}} \bibnamefont{and}
  \bibinfo{author}{\bibfnamefont{P.}~\bibnamefont{Horodecki}},
  \bibinfo{journal}{Physical Review A} \textbf{\bibinfo{volume}{61}},
  \bibinfo{pages}{040301} (\bibinfo{year}{2000}).

\bibitem[{\citenamefont{Saks and Zygmund}(1952)}]{saks1952analytic}
\bibinfo{author}{\bibfnamefont{S.}~\bibnamefont{Saks}} \bibnamefont{and}
  \bibinfo{author}{\bibfnamefont{A.}~\bibnamefont{Zygmund}},
  \emph{\bibinfo{title}{Analytic functions}}, vol.~\bibinfo{volume}{28}
  (\bibinfo{publisher}{Monografie Matematyczne}, \bibinfo{year}{1952}).

\bibitem[{\citenamefont{Zavatta et~al.}(2004)\citenamefont{Zavatta, Viciani,
  and Bellini}}]{zavatta2004quantum}
\bibinfo{author}{\bibfnamefont{A.}~\bibnamefont{Zavatta}},
  \bibinfo{author}{\bibfnamefont{S.}~\bibnamefont{Viciani}}, \bibnamefont{and}
  \bibinfo{author}{\bibfnamefont{M.}~\bibnamefont{Bellini}},
  \bibinfo{journal}{science} \textbf{\bibinfo{volume}{306}},
  \bibinfo{pages}{660} (\bibinfo{year}{2004}).

\bibitem[{\citenamefont{Marco and Alessandro}(2010)}]{marco2010manipulating}
\bibinfo{author}{\bibfnamefont{B.}~\bibnamefont{Marco}} \bibnamefont{and}
  \bibinfo{author}{\bibfnamefont{Z.}~\bibnamefont{Alessandro}}, in
  \emph{\bibinfo{booktitle}{Progress in Optics}}
  (\bibinfo{publisher}{Elsevier}, \bibinfo{year}{2010}),
  vol.~\bibinfo{volume}{55}, pp. \bibinfo{pages}{41--83}.

\bibitem[{\citenamefont{Walschaers et~al.}(2018)\citenamefont{Walschaers,
  Sarkar, Parigi, and Treps}}]{walschaers2018tailoring}
\bibinfo{author}{\bibfnamefont{M.}~\bibnamefont{Walschaers}},
  \bibinfo{author}{\bibfnamefont{S.}~\bibnamefont{Sarkar}},
  \bibinfo{author}{\bibfnamefont{V.}~\bibnamefont{Parigi}}, \bibnamefont{and}
  \bibinfo{author}{\bibfnamefont{N.}~\bibnamefont{Treps}},
  \bibinfo{journal}{Physical review letters} \textbf{\bibinfo{volume}{121}},
  \bibinfo{pages}{220501} (\bibinfo{year}{2018}).

\bibitem[{\citenamefont{Yadin et~al.}(2018)\citenamefont{Yadin, Binder,
  Thompson, Narasimhachar, Gu, and Kim}}]{yadin2018operational}
\bibinfo{author}{\bibfnamefont{B.}~\bibnamefont{Yadin}},
  \bibinfo{author}{\bibfnamefont{F.~C.} \bibnamefont{Binder}},
  \bibinfo{author}{\bibfnamefont{J.}~\bibnamefont{Thompson}},
  \bibinfo{author}{\bibfnamefont{V.}~\bibnamefont{Narasimhachar}},
  \bibinfo{author}{\bibfnamefont{M.}~\bibnamefont{Gu}}, \bibnamefont{and}
  \bibinfo{author}{\bibfnamefont{M.~S.} \bibnamefont{Kim}},
  \bibinfo{journal}{Physical Review X} \textbf{\bibinfo{volume}{8}},
  \bibinfo{pages}{041038} (\bibinfo{year}{2018}).

\bibitem[{\citenamefont{Menzies and Filip}(2009)}]{menzies2009gaussian}
\bibinfo{author}{\bibfnamefont{D.}~\bibnamefont{Menzies}} \bibnamefont{and}
  \bibinfo{author}{\bibfnamefont{R.}~\bibnamefont{Filip}},
  \bibinfo{journal}{Physical Review A} \textbf{\bibinfo{volume}{79}},
  \bibinfo{pages}{012313} (\bibinfo{year}{2009}).

\bibitem[{\citenamefont{Straka et~al.}(2018)\citenamefont{Straka, Lachman,
  Hlou{\v{s}}ek, Mikov{\'a}, Mi{\v{c}}uda, Je{\v{z}}ek, and
  Filip}}]{straka2018quantum}
\bibinfo{author}{\bibfnamefont{I.}~\bibnamefont{Straka}},
  \bibinfo{author}{\bibfnamefont{L.}~\bibnamefont{Lachman}},
  \bibinfo{author}{\bibfnamefont{J.}~\bibnamefont{Hlou{\v{s}}ek}},
  \bibinfo{author}{\bibfnamefont{M.}~\bibnamefont{Mikov{\'a}}},
  \bibinfo{author}{\bibfnamefont{M.}~\bibnamefont{Mi{\v{c}}uda}},
  \bibinfo{author}{\bibfnamefont{M.}~\bibnamefont{Je{\v{z}}ek}},
  \bibnamefont{and} \bibinfo{author}{\bibfnamefont{R.}~\bibnamefont{Filip}},
  \bibinfo{journal}{npj Quantum Information} \textbf{\bibinfo{volume}{4}},
  \bibinfo{pages}{4} (\bibinfo{year}{2018}).

\bibitem[{\citenamefont{Wang et~al.}(2013)\citenamefont{Wang, Yuan, and
  Xu}}]{wang2013conditional}
\bibinfo{author}{\bibfnamefont{S.}~\bibnamefont{Wang}},
  \bibinfo{author}{\bibfnamefont{H.-c.} \bibnamefont{Yuan}}, \bibnamefont{and}
  \bibinfo{author}{\bibfnamefont{X.-f.} \bibnamefont{Xu}},
  \bibinfo{journal}{The European Physical Journal D}
  \textbf{\bibinfo{volume}{67}}, \bibinfo{pages}{102} (\bibinfo{year}{2013}).

\bibitem[{\citenamefont{Su et~al.}(2019)\citenamefont{Su, Myers, and
  Sabapathy}}]{su2019conversion}
\bibinfo{author}{\bibfnamefont{D.}~\bibnamefont{Su}},
  \bibinfo{author}{\bibfnamefont{C.~R.} \bibnamefont{Myers}}, \bibnamefont{and}
  \bibinfo{author}{\bibfnamefont{K.~K.} \bibnamefont{Sabapathy}},
  \bibinfo{journal}{arXiv preprint arXiv:1902.02323}  (\bibinfo{year}{2019}).

\bibitem[{\citenamefont{Nielsen and Chuang}(2011)}]{NielsenChuang}
\bibinfo{author}{\bibfnamefont{M.~A.} \bibnamefont{Nielsen}} \bibnamefont{and}
  \bibinfo{author}{\bibfnamefont{I.~L.} \bibnamefont{Chuang}},
  \emph{\bibinfo{title}{Quantum Computation and Quantum Information: 10th
  Anniversary Edition}} (\bibinfo{publisher}{Cambridge University Press},
  \bibinfo{address}{New York, NY, USA}, \bibinfo{year}{2011}),
  \bibinfo{edition}{10th} ed.

\bibitem[{\citenamefont{Stenholm}(1992)}]{stenholm1992simultaneous}
\bibinfo{author}{\bibfnamefont{S.}~\bibnamefont{Stenholm}},
  \bibinfo{journal}{annals of physics} \textbf{\bibinfo{volume}{218}},
  \bibinfo{pages}{233} (\bibinfo{year}{1992}).

\bibitem[{\citenamefont{Stein and Shakarchi}(2010)}]{stein2010complex}
\bibinfo{author}{\bibfnamefont{E.~M.} \bibnamefont{Stein}} \bibnamefont{and}
  \bibinfo{author}{\bibfnamefont{R.}~\bibnamefont{Shakarchi}},
  \emph{\bibinfo{title}{Complex analysis}}, vol.~\bibinfo{volume}{2}
  (\bibinfo{publisher}{Princeton University Press}, \bibinfo{year}{2010}).

\bibitem[{\citenamefont{Wyss}(2017)}]{wyss2017two}
\bibinfo{author}{\bibfnamefont{W.}~\bibnamefont{Wyss}}, \bibinfo{journal}{arXiv
  preprint arXiv:1707.03861}  (\bibinfo{year}{2017}).

\bibitem[{\citenamefont{Abramowitz and Stegun}(1965)}]{abramowitz1965handbook}
\bibinfo{author}{\bibfnamefont{M.}~\bibnamefont{Abramowitz}} \bibnamefont{and}
  \bibinfo{author}{\bibfnamefont{I.~A.} \bibnamefont{Stegun}},
  \emph{\bibinfo{title}{Handbook of mathematical functions: with formulas,
  graphs, and mathematical tables}}, vol.~\bibinfo{volume}{55}
  (\bibinfo{publisher}{Courier Corporation}, \bibinfo{year}{1965}).

\end{thebibliography}


\end{document}